# Technical Dimensions of Programming Systems

Joel Jakubovic[a], Jonathan Edwards[b], and Tomas Petricek[a,c]

a   University of Kent, Canterbury, UK
b   Independent
c   Charles University, Prague, Czechia

**Abstract**   Programming requires much more than just writing code in a programming language. It is usually done in the context of a stateful environment, by interacting with a system through a graphical user interface. Yet, this wide space of possibilities lacks a common structure for navigation. Work on programming systems fails to form a coherent body of research, making it hard to improve on past work and advance the state of the art.

In computer science, much has been said and done to allow comparison of *programming languages,* yet no similar theory exists for *programming systems;* we believe that programming systems deserve a theory too.

We present a framework of *technical dimensions* which capture the underlying characteristics of programming systems and provide a means for conceptualizing and comparing them.

We identify technical dimensions by examining past influential programming systems and reviewing their design principles, technical capabilities, and styles of user interaction. Technical dimensions capture characteristics that may be studied, compared and advanced independently. This makes it possible to talk about programming systems in a way that can be shared and constructively debated rather than relying solely on personal impressions.

Our framework is derived using a qualitative analysis of past programming systems. We outline two concrete ways of using our framework. First, we show how it can analyze a recently developed novel programming system. Then, we use it to identify an interesting unexplored point in the design space of programming systems.

Much research effort focuses on building programming systems that are easier to use, accessible to non-experts, moldable and/or powerful, but such efforts are disconnected. They are informal, guided by the personal vision of their authors and thus are only evaluable and comparable on the basis of individual experience using them. By providing foundations for more systematic research, we can help programming systems researchers to stand, at last, on the shoulders of giants.

## The Art, Science, and Engineering of Programming

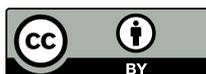

> A systematic presentation removes ideas from the ground that made them grow and arranges them in an artificial pattern.

*The Tyranny of Science*
PAUL FEYERABEND

> Irony is said to allow the artist to continue his creative production while immersed in a sociocultural context of which he is critical.

*Irony; or, the Self-Critical Opacity of Postmodernist Architecture*
EMMANUEL PETIT

## 1 Introduction

Many forms of software have been developed to enable programming. The classic form consists of a *programming language*, a text editor to enter source code, and a compiler to turn it into an executable program. Instances of this form are differentiated by the syntax and semantics of the language, along with the implementation techniques in the compiler or runtime environment. Since the advent of graphical user interfaces (GUIs), programming languages can be found embedded within graphical environments that increasingly define how programmers work with the language—for instance, by directly supporting debugging or refactoring. Beyond this, the rise of GUIs also permits diverse visual forms of programming, including visual languages and GUI-based end-user programming tools.

This paper advocates a shift of attention from *programming languages* to the more general notion of "software that enables programming"—in other words, *programming systems*.

**Definition** (Programming System). *A* programming system *is an integrated and complete set of tools sufficient for creating, modifying, and executing programs. These will include notations for structuring programs and data, facilities for running and debugging programs, and interfaces for performing all of these tasks. Facilities for testing, analysis, packaging, or version control may also be present. Notations include programming languages and interfaces include text editors, but are not limited to these.*

This notion covers classic programming languages together with their editors, debuggers, compilers, and other tools. Yet it is intentionally broad enough to accommodate image-based programming environments like Smalltalk, operating systems like UNIX, and hypermedia authoring systems like Hypercard, in addition to various other examples we will mention.

### 1.1 The Problem: No Systematic Framework for Systems

There is a growing interest in broader forms of *programming systems*, both in the programming research community and in industry. Researchers are studying topics such as *programming experience* and *live programming* that require considering not just the *language*, but further aspects of a given system. At the same time, commercial companies are building new programming environments like Replit [78] or low-code tools like Dark [56] and Glide [30]. Yet, such topics remain at the sidelines of mainstream programming research. While *programming languages* are a well-established concept, analysed and compared in a common vocabulary, no similar foundation exists for the wider range of *programming systems*.

The academic research on programming suffers from this lack of common vocabulary. While we may thoroughly assess programming *languages*, as soon as we add interaction or graphics into the picture, evaluation beyond subjective "coolness" becomes fraught with difficulty.[1] Moreover, when designing new systems, inspiration is often drawn from the same few standalone sources of ideas. These might be influential past systems like Smalltalk, programmable end-user applications like spreadsheets, or motivational illustrations like those of Bret Victor [90].

Instead of forming a solid body of work, the ideas that emerge are difficult to relate to each other. The research methods used to study programming systems lack the rigorous structure of programming language research methods. They tend to rely on singleton examples, which demonstrate their author's ideas, but are inadequate methods for comparing new ideas with the work of others. This makes it hard to build on top and thereby advance the state of the art.

Studying *programming systems* is not merely about taking a programming language and looking at the tools that surround it. It presents a *paradigm shift* to a perspective that is, at least partly, *incommensurable* with that of languages. When studying programming languages, everything that matters is in the program code; when studying programming systems, everything that matters is in the *interaction* between the programmer and the system. As documented by Gabriel [28], looking at a *system* from a *language* perspective makes it impossible to think about concepts that arise from interaction with a system, but are not reflected in the language. Thus, we must proceed with some caution. As we will see, when we talk about Lisp as a programming system, we mean something very different from a parenthesis-heavy programming language!

---

[1] The same difficulty in the context of user interface systems has been analyzed by Olsen [68]. Interesting future work would be a detailed analysis of publications on programming systems to understand this issue in depth. One notable characteristic is that publications tend to present (parts of) new systems. This is the case for 5/6 and 6/7 papers in the LIVE 2020 and 2021 workshops respectively [39, 37]. In contrast, publications in the field of programming *languages* often address specific issues of interest to a greater number of languages.

## 1.2 Contributions

We propose a common language as an initial step towards a more progressive research on programming systems. Our set of *technical dimensions* seeks to break down the holistic view of systems along various specific "axes". The dimensions identify a range of possible design choices, characterized by two extreme points in the design space. They are not quantitative, but they allow comparison by locating systems on a common axis. We do not intend for the extreme points to represent "good" or "bad" designs; we expect any position to be a result of design trade-offs. At this early stage in the life of such a framework, we encourage agreement on descriptions of systems first in order to settle any normative judgements later.

The set of dimensions can be understood as a map of the design space of programming systems (Figure 1). Past and present systems will serve as landmarks, and with enough of them, we may reveal unexplored or overlooked possibilities. So far, the field has not been able to establish a virtuous cycle of feedback; it is hard for practitioners to situate their work in the context of others' so that subsequent work can improve on it. Our aim is to provide foundations for the study of programming systems that would allow such development.

This paper is intended as a reference on the current state of the technical dimensions framework and it is meant to be *used* rather than *read*. We present the dimensions in detail, but encourage the reader to skim through the details on the first read. Subsequently, we suggest revisiting dimensions which seem relevant to a concrete system known to the reader. The main contributions of this paper are organized as follows:

1. In Section 3, we characterize what a programming system is and review landmark programming systems of the past that are used as examples throughout this paper, as well as to delineate our notion of a programming system.

2. We present the technical dimensions in detail, organised into related clusters: *interaction*, *notation*, *conceptual structure*, *customizability*, *complexity*, *errors*, and *adoptability*. For each dimension, we give examples that illustrate the range of values along its axis. We intend this as a reference to be used as needed rather than something to be read from start to finish, so we recommend skimming the catalogue on the first reading.

3. In Section 5, we sketch two ways of using the technical dimensions framework. In Section 5.1, we use it to evaluate a recent interesting programming system; in Section 5.2, we use it to identify an unexplored point in the design space and envision a potential novel programming system.

## 2   Related Work

While we do have new ideas to propose, part of our contribution is integrating a wide range of *existing* concepts under a common umbrella. This work is spread out across

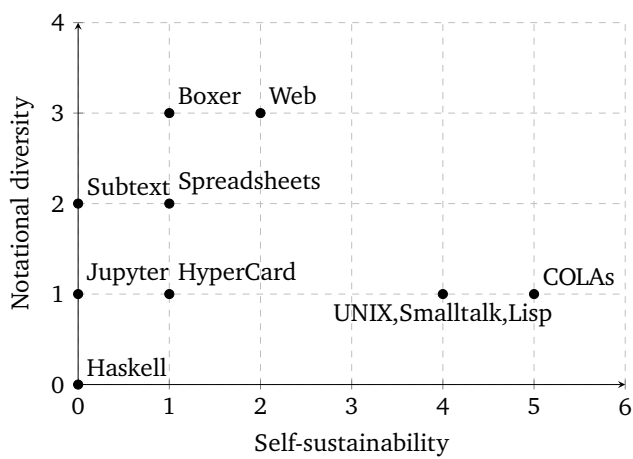

■ **Figure 1** One 2-dimensional slice of the space of possible systems, to be examined in more detail in Section 5.2. The numerical scores on each axis are generated systematically by a method described in Appendix A. While these results are plausible, they are not definitive as the method could be developed a lot further in future work (see Section A.3).

different domains, but each part connects to programming systems or focuses on a specific characteristic they may have.

**From languages to systems.** Our approach lies between a narrow focus on programming languages and a broad focus on programming as a socio-political and cultural subject. Our concept of a programming system is technical in scope, although we acknowledge the technical side often has important social implications as in the case of the "Adoptability" dimension (Section 4.7). This contrasts with the more socio-political focus found in Tchernavskij [88] or in software studies [25]. It overlaps with Kell's conceptualization of UNIX, Smalltalk, and Operating Systems generally [45], and we have ensured that UNIX has a place in our framework.

The distinction between more narrow *programming languages* and broader *programming systems* is more subtle. Richard Gabriel noted an invisible paradigm shift from the study of "systems" to the study of "languages" in computer science [28], and this observation informs our distinction here. One consequence of the change is that a *language* is often formally specified apart from any specific implementations, while *systems* resist formal specification and are often *defined by* an implementation. We recognize programming language implementations as a *small region* of the space of possible systems, at least as far as interaction and notations might go. Hence we refer to the *interactive programming system* aspects of languages, such as text editing and command-line workflow.

**Programming systems research.** There is renewed interest in programming systems in the form of recent non-traditional programming tools:

- Computational notebooks such as Jupyter [51] facilitate data analysis by combining code snippets with text and visual output. They are backed by stateful "kernels" and used interactively.
- "Low code" end-user programming systems allow application development (mostly) through a GUI. One example is Coda [12], which combines tables, formulas, and scripts to enable non-technical people to build "applications as documents".
- Domain-specific programming systems such as Dark [56], which claims a "holistic" programming experience for cloud API services. This includes a language, a direct manipulation editor, and near-instantaneous building and deployment.
- Even for general purpose programming with conventional tools, systems like Replit [78] have demonstrated the benefits of integrating all needed languages, tools, and user interfaces into a seamless experience, available from the browser, that requires no setup.

Research that follows the programming systems perspective can be found in a number of research venues. Those include Human-Computer Interaction conferences such as UIST[2] and VL/HCC.[3] However, work in those often emphasizes the user experience over technical description. Programming systems are often presented in workshops such as LIVE and PX.[4] However, work in those venues is often limited to the authors' individual perspectives and suffers from the aforementioned difficulty of comparing to other systems.

Concrete examples of systems appear throughout the paper. Recent systems which motivated some of our dimensions include Subtext [17], which combines code with its live execution in a single editable representation; Sketch-n-sketch [38], which can synthesize code by direct manipulation of its outputs; Hazel [69], a live functional programming environment with typed holes to enable execution of incomplete or ill-typed programs; and Webstrates [50], which extends Web pages with real-time sharing of state.

**Already-known characteristics.** There are several existing projects identifying characteristics of programming systems. Some revolve around a single one, such as levels of liveness [87], or plurality and communicativity [46]. Others propose an entire collection. *Memory Models of Programming Languages* [81] identifies the "everything is an X" metaphors underlying many programming languages; the *Design Principles of Smalltalk* [42] documents the philosophical goals and dicta used in the design of Smalltalk; the "Gang of Four" *Design Patterns* [29] catalogues specific implementation tactics; and the *Cognitive Dimensions of Notation* [34] identifies a common vocabulary for software's *notational surface* and for identifying their trade-offs.

The latter two directly influence our proposal. Firstly, the Cognitive Dimensions are a set of qualitative properties which can be used to analyze *notations*. We are extending this approach to the "rest" of a system, beyond its notation, with *Technical* Dimensions.

---

[2] ACM Symposium on User Interface Software and Technology.
[3] IEEE Symposium on Visual Languages and Human-Centric Computing.
[4] Programming eXperience.

Secondly, our individual dimensions naturally fall under larger *clusters* that we present in a regular format, similar to the presentation of the classic Design Patterns. As for characteristics identified by others, part of our contribution is to integrate them under a common umbrella: the existing concepts of liveness, pluralism, and uniformity metaphors ("everything is an X") become dimensions in our framework.

**Methodology.** We follow the attitude of *Evaluating Programming Systems* [18] in distinguishing our work from HCI methods and empirical evaluation. We are generally concerned with characteristics that are not obviously amenable to statistical analysis (e.g. mining software repositories) or experimental methods like controlled user studies, so numerical quantities are generally not featured.

Similar development seems to be taking place in HCI research focused on user interfaces. The UIST guidelines [55] instruct authors to evaluate system contributions holistically, and the community has developed heuristics for such evaluation, such as *Evaluating User Interface Systems Research* [68]. Our set of dimensions offers similar heuristics for identifying interesting aspects of programming systems, though they focus more on underlying technical properties than the surface interface.

Finally, we believe that the aforementioned paradigm shift from programming systems to programming languages has hidden many ideas about programming that are worthwhile recovering and developing further [74]. Thus our approach is related to the idea of *complementary science* developed by Chang [9] in the context of history and philosophy of science. Chang argues that even in disciplines like physics, superseded or falsified theories may still contain interesting ideas worth documenting. In the field of programming, where past systems are discarded for many reasons besides empirical failure, Chang's *complementary science* approach seems particularly suitable.

**Programming systems deserve a theory too!** In short, while there is a theory for programming languages, programming *systems* deserve a theory too. It should apply from the small scale of language implementations to the vast scale of operating systems. It should be possible to analyse the common and unique features of different systems, to reveal new possibilities, and to build on past work in an effective manner. In Kuhnian terms [54], it should enable a body of "normal science": filling in the map of the space of possible systems (Figure 1), thereby forming a knowledge repository for future designers.

## 3 Programming Systems

We introduce the notion of a *programming system* through a number of example systems. We draw them from three broad reference classes:

- Software ecosystems built around a text-based programming *language*. They consist of a set of tools such as compilers, debuggers, and profilers. These tools may exist as separate command-line programs, or within an Integrated Development Environment.

- Those that resemble an *operating system* (OS) in that they structure the execution environment and encompass the resources of an entire machine (physical or virtual). They provide a common interface for communication, both between the user and the computer, and between programs themselves.
- Programmable *applications*, typically optimized for a specific domain, offering a limited degree of programmability which may be increased with newer versions.

It must be noted that our selection of systems is not meant to be exhaustive; there will be many past and present systems that we are not aware of or do not know much about, and we obviously cannot cover programming systems that have not been created yet. With that caveat, we will proceed to detail some systems under the above grouping. This will provide an intuition for the notion of a programming system and establish a collection of go-to examples for the rest of the paper.

## 3.1 Systems Based Around Languages

Text-based programming languages sit within programming systems whose boundaries are not explicitly defined. To speak of a programming system, we need to consider a language with, at minimum, an editor and a compiler or interpreter. However, the exact boundaries are a design choice that significantly affects our analysis.

**Java with the Eclipse ecosystem.**   Java [32] cannot be viewed as a programming system on its own, but it is one if we consider it as embedded in an ecosystem of tools. There are multiple ways to delineate this, resulting in different analyses. A minimalistic programming system would consist of a text editor to write Java code and a command line compiler. A more realistic system is Java as embedded in the Eclipse IDE [14]. The programming systems view allows us to see all there is beyond the textual code. In the case of Eclipse, this includes the debugger, refactoring tools, testing and modelling tools, GUI designers, and so on. This delineation yields a programming system that is powerful and convenient, but has a large number of concepts and secondary notations.

**Haskell tools ecosystem.**   Haskell is an even more language-focused programming system. It is used through the command-line *GHC* compiler [59] and *GHCi* REPL, alongside a text editor that provides features like syntax highlighting and auto-completion. Any editor that supports the Language Server Protocol [63] will suffice to complete the programming system.

Haskell is mathematically rooted and relies on mathematical intuition for understanding many of its concepts. This background is also reflected in the notations it uses. In addition to the concrete language syntax for writing code, the ecosystem also uses an informal mathematical notation for writing about Haskell (e.g. in academic papers or on the whiteboard). This provides an additional tool for manipulating Haskell programs. Experiments on paper can provide a kind of rapid feedback that other systems may provide through live programming.

**From REPLs to notebooks.** A different kind of developer ecosystem that evolved around a programming language is the Jupyter notebook platform [51]. In Jupyter, data scientists write scripts divided into notebook cells, execute them interactively and see the resulting data and visualizations directly in the notebook itself. This brings together the REPL, which dates back to conversational implementations of Lisp in the 1960s, with literate programming [52] used in the late 1980s in Mathematica 1.0 [93].

As a programming system, Jupyter has a number of interesting characteristics. The primary outcome of programming is the notebook itself, rather than a separate application to be compiled and run. The code lives in a document format, interleaved with other notations. Code is written in small parts that are executed quickly, offering the user more rapid feedback than in conventional programming. A notebook can be seen as a trace of how the result has been obtained, yet one often problematic feature of notebooks is that some allow the user to run code blocks out-of-order. The code manipulates mutable state that exists in a "kernel" running in the background. Thus, retracing one's steps in a notebook is more subtle than in, say, Common Lisp [85], where the dribble function would directly record the user's session to a file.

### 3.2 OS-like Programming Systems

"OS-likes" begin from the 1960s when it became possible to interact one-on-one with a computer. First, time-sharing systems enabled interactive shared use of a computer via a teletype; smaller computers such as the PDP-1 and PDP-8 provided similar direct interaction, while 1970s workstations such as the Alto and Lisp Machines added graphical displays and mouse input. These *OS-like* systems stand out as having the totalising scope of *operating systems*, whether or not they are ordinarily seen as taking this role.

**MacLisp and Interlisp.** LISP 1.5 [60] arrived before the rise of interactive computers, but the existence of an interpreter and the absence of declarations made it natural to use Lisp interactively, with the first such implementations appearing in the early 1960s. Two branches of the Lisp family [86], MacLisp and the later Interlisp, embraced the interactive "conversational" way of working, first through a teletype and later using the screen and keyboard.

Both MacLisp and Interlisp adopted the idea of *persistent address space*. Both program code and program state were preserved when powering off the system, and could be accessed and modified interactively as well as programmatically using the *same means*. Lisp Machines embraced the idea that the machine runs continually and saves the state to disk when needed. Today, this is widely seen in cloud-based services like Google Docs and online IDEs. Another idea pioneered in MacLisp and Interlisp was the use of *structure editors*. These let programmers work with Lisp data structures not as sequences of characters, but as nested lists. In Interlisp, the programmer would use commands such as *P to print the current expression, or *(2 (X Y)) to replace its second element with the argument (X Y). The PILOT system [89] offered even more sophisticated conversational features. For typographical errors and other slips, it would offer an automatic fix for the user to interactively accept, modifying the program in memory and resuming execution.

**Smalltalk.**   Smalltalk appeared in the 1970s with a distinct ambition of providing "dynamic media which can be used by human beings of all ages" [43]. The authors saw computers as *meta-media* that could become a range of other media for education, discourse, creative arts, simulation and other applications not yet invented. Smalltalk was designed for single-user workstations with a graphical display, and pioneered this display not just for applications but also for programming itself. In Smalltalk 72, one wrote code in the bottom half of the screen using a structure editor controlled by a mouse, and menus to edit definitions. In Smalltalk-76 and later, this had switched to text editing embedded in a *class browser* for navigating through classes and their methods.

Similarly to Lisp systems, Smalltalk adopts the persistent address space model of programming where all objects remain in memory, but based on *objects* and *message passing* rather than *lists*. Any changes made to the system state by programming or execution are preserved when the computer is turned off. Lastly, the fact that much of the Smalltalk environment is implemented in itself makes it possible to extensively modify the system from within.

We include Lisp and Smalltalk in the OS-likes because they function as operating systems in many ways. On specialized machines, like the Xerox Alto and Lisp machines, the user started their machine directly in the Lisp or Smalltalk environment and was able to do everything they needed from *within* the system. Nowadays, however, this experience is associated with UNIX and its descendants on a vast range of commodity machines.

**UNIX.**   UNIX illustrates the fact that many aspects of programming systems are shaped by their intended target audience. Built for computer hackers [57], its abstractions and interface are close to the machine. Although historically linked to the C language, UNIX developed a language-agnostic set of abstractions that make it possible to use multiple programming languages in a single system. While everything is an object in Smalltalk, the ontology of the UNIX system consists of files, memory, executable programs, and running processes. Note the explicit "stage" distinction here: UNIX distinguishes between volatile *memory* structures, which are lost when the system is shut down, and non-volatile *disk* structures that are preserved. This distinction between types of memory is considered, by Lisp and Smalltalk, to be an implementation detail to be abstracted over by their persistent address space. Still, this did not prevent the UNIX ontology from supporting a pluralistic ecosystem of different languages and tools.

**Early and modern Web.**   The Web evolved [1] from a system for sharing and organizing information to a *programming system*. Today, it consists of a wide range of server-side programming tools, JavaScript and languages that compile to it, and notations like HTML and CSS. As a programming system, the "modern 2020s web" is reasonably distinct from the "early 1990s web". In the early web, JavaScript code was distributed in a form that made it easy to copy and re-use existing scripts, which led to enthusiastic adoption by non-experts—recalling the birth of microcomputers like Commodore 64 with BASIC a decade earlier.

In the "modern web", multiple programming languages treat JavaScript as a compilation target, and JavaScript is also used as a language on the server-side. This web is no longer simple enough to encourage copy-and-paste remixing of code from different sites. However, it does come with advanced developer tools that provide functionality resembling early interactive programming systems like Lisp and Smalltalk. The *Document Object Model (DOM)* structure created by a web page is transparent, accessible to the user and modifiable through the built-in browser inspector tools. Third-party code to modify the DOM can be injected via extensions. The DOM almost resembles the tree/graph model of Smalltalk and Lisp images, lacking the key persistence property. This limitation, however, is being addressed by Webstrates [50].

### 3.3 Application-Focused Systems

The previously discussed programming systems were either universal, not focusing on any particular kind of application, or targeted at broad fields, such as Artificial Intelligence and symbolic data manipulation in Lisp's case. In contrast, the following examples focus on more narrow kinds of applications that need to be built. Many support programming based on rich interactions with specialized visual and textual notations.

**Spreadsheets.** The first spreadsheets became available in 1979 in VisiCalc [33, 94] and helped analysts perform budget calculations. As programming systems, spreadsheets are notable for their programming substrate (a two-dimensional grid) and evaluation model (automatic re-evaluation). The programmability of spreadsheets developed over time, acquiring features that made them into powerful programming systems in a way VisiCalc was not. The final step was the 1993 inclusion of *macros* in Excel, later further extended with *Visual Basic for Applications*.

**Graphical "languages".** Efforts to support programming without relying on textual code can only be called "languages" in a metaphorical sense. In these programming systems, programs are made out of graphical structures as in LabView [53] or Programming-By-Example [58].

**HyperCard.** While spreadsheets were designed to solve problems in a specific application area, HyperCard [62] was designed around a particular application format. Programs are "stacks of cards" containing multimedia components and controls such as buttons. These controls can be programmed with pre-defined operations like "navigate to another card", or via the HyperTalk scripting language for anything more sophisticated.

As a programming system, HyperCard is interesting for a couple of reasons. It effectively combines visual and textual notation. Programs appear the same way during editing as they do during execution. Most notably, HyperCard supports gradual progression from the "user" role to "developer": a user may first use stacks, then go on to edit the visual aspects or choose pre-defined logic until, eventually, they learn to program in HyperTalk.

## 4  Technical Dimensions Catalogue

Here, we present our proposed technical dimensions in great detail. Please note that our intention is to provide a *reference* to be looked up and *used* as needed, not something that should be read from start to finish. Therefore, we recommend skimming through this for anything particularly interesting before proceeding to Section 5. There, we will reference several dimensions in the context of a specific example, at which point it may be helpful to come back for more detail. For a quick overview, we include summary tables 1, 2, and 3, though they may make more sense after reading the relevant sections.

We present the dimensions grouped under *clusters*. These may be regarded as "topics of interest" or "areas of inquiry" when studying a given system, grouping together related dimensions against which to measure it.

Each cluster is named and opens with a boxed *summary*, followed by a longer *discussion*, and closes with a list of any *relations* to other clusters along with any *references* if applicable. Within the main discussion, individual *dimensions* are listed. Sometimes, a particular value along a dimension (or a combination of values along several dimensions) can be recognized as a familiar pattern—perhaps with a name already established in the literature. These are marked as *examples*. Finally, interspersed discussion that is neither a *dimension* nor an *example* is introduced as a *remark*.

We reiterate that we do not expect the proposed catalogue to be exhaustive for all aspects of programming systems, past and future. We welcome follow-up work expanding or challenging the aspects we have focused on here.

### 4.1  Interaction

**How do users manifest their ideas, evaluate the result, and generate new ideas in response?**

An essential aspect of programming systems is how the user interacts with them when creating programs. Take the standard form of statically typed, compiled languages with straightforward library linking: here, programmers write their code in a text editor, invoke the compiler, and read through error messages they get. After fixing the code to pass compilation, a similar process might happen with runtime errors.

Other forms are yet possible. On the one hand, some typical interactions like compilation or execution of a program may not be perceptible at all. On the other hand, the system may provide various interfaces to support the plethora of other interactions that are often important in programming, such as looking up documentation, managing dependencies, refactoring or pair programming.

■ **Table 1**  Technical Dimensions cheat sheet.

| Dimension (CLUSTER) | Summary | Range of key examples |
|---|---|---|
| INTERACTION | How do users manifest their ideas, evaluate the result, and generate new ideas in response? | |
| Feedback Loops | How wide are the various gulfs of *execution* and *evaluation* and how are they related? | Immediate Feedback (short) vs. batch mode (long) gulf of evaluation |
| Modes of Interaction | Which sets of feedback loops only occur together? | Setup vs. editing vs. debugging |
| Abstraction Construction | How do we go from abstractions to concrete examples and vice versa? | Programming by Example vs. first principles |
| NOTATION | How are the different textual / visual programming notations related? | |
| Notational Structure | What notations are used to program the system and how are they related? | Notations overlap and need sync vs. complement each other |
| Surface / Internal Notations | What is the connection between what a user sees and what a computer program sees? | Sequence Editing vs. Rendering, Structure Editing vs. Recovery |
| Primary / Secondary Notations | Is one notation more important than others? | Secondary build scripts vs. visual editor and code on equal footing in Flash |
| Expression Geography | Do similar expressions encode similar programs? | Concise yet error-prone vs. explicit yet verbose |
| Uniformity of Notations | Does the notation use a small or a large number of basic concepts? | Lisp S-expressions vs. English-like textual notations |

■ **Table 2** Technical Dimensions cheat sheet (continued).

| Dimension (CLUSTER) | Summary | Range of key examples |
|---|---|---|
| CONCEPTUAL STRUCTURE | How is meaning constructed? How are internal and external incentives balanced? | |
| Conceptual Integrity vs. Openness | Does the system present as elegantly *designed* or pragmatically *improvised*? | Integrity (Everything is a X) vs. openness (compatible mixtures) |
| Composability | What are the primitives? How can they be combined to achieve novel behaviors? | Sequence, selection, repetition, function abstraction, recursion, logical connectives |
| Convenience | Which wheels do users not need to reinvent? | Small vs. expansive standard libraries |
| Commonality | How much is *common structure* explicitly marked as such? | Common structure is redundantly flattened vs. factored out |
| CUSTOMIZABILITY | Once a program exists in the system, how can it be extended and modified? | |
| Staging of Customization | Must we customize *running* programs differently to *inert* ones? Do these changes last beyond termination? | Source code vs. config files, Developer Tools tab, auto image-based persistence, scripting language |
| Externalizability / Additive Authoring | Which portions of the system's state can be referenced and transferred to/from it? How far can the system's behavior be changed by *adding* expressions? | State is: (*i*) private and hidden; (*ii*) externalizable and overwriteable; (*iii*) externalized and *overridable* by addition |
| Self-Sustainability | How far can the system's behavior be changed from within? | None (rely on external tools) vs. self-sufficient (contains everything needed) |

■ **Table 3** Technical Dimensions cheat sheet (continued).

| Dimension (CLUSTER) | Summary | Range of key examples |
|---|---|---|
| COMPLEXITY | How does the system structure complexity and what level of detail is required? | |
| Factoring of Complexity | What programming details are hidden in reusable components and how? | Domain-specific (more hiding) vs. general-purpose (less hiding) |
| Level of Automation | What part of program logic does not need to be explicitly specified? | Garbage collection (low-tech) vs. Prolog engine (hi-tech) |
| ERRORS | What does the system consider to be an *error*? How are they prevented and handled? | |
| Error Detection | What errors can be detected in which feedback loops, and how? | Human inspection in live coding vs. partial automation in static typing |
| Error Response | How does the system respond when an error is detected? | Does it stop, recover automatically, ignore the error or ask the user how to continue? |
| ADOPTABILITY | How does the system facilitate or obstruct adoption by both individuals and communities? | |
| Learnability | What is the attitude towards the *learning curve* and what is the target audience? | HyperCard for the general public vs. FORTRAN for scientists |
| Sociability | What are the social and economic factors that make the system the way it is? | Cathedral vs. Bazaar model |

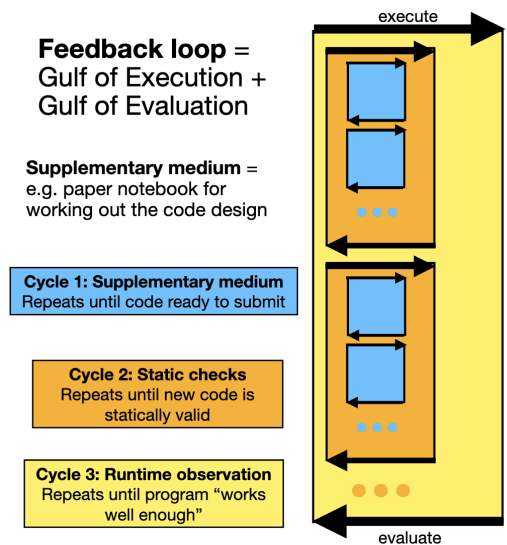

**Feedback loop** =
Gulf of Execution +
Gulf of Evaluation

**Supplementary medium** =
e.g. paper notebook for
working out the code design

**Cycle 1: Supplementary medium**
Repeats until code ready to submit

**Cycle 2: Static checks**
Repeats until new code is
statically valid

**Cycle 3: Runtime observation**
Repeats until program "works
well enough"

execute

evaluate

■ **Figure 2**  The nested feedback loops of a statically-checked programming language.

We focus on the interactions where programmer interacts with the system to construct a program with a desired behavior. To analyze those, we use the concepts of *gulf of execution* and *gulf of evaluation* from *The Design of Everyday Things* [67].

### 4.1.1 Dimension: Feedback Loops

In using a system, one first has some idea and attempts to make it exist in the software; the gap between the user's goal and the means to execute the goal is known as the *gulf of execution*. Then, one compares the result actually achieved to the original goal in mind; this crosses the *gulf of evaluation*. These two activities comprise the *feedback loop* through which a user gradually realises their desires in the imagination, or refines those desires to find out "what they actually want".

A system must contain at least one such feedback loop, but may contain several at different levels or specialized to certain domains. For each of them, we can separate the gulf of execution and evaluation as independent legs of the journey, with possibly different manners and speeds of crossing them.

For example, we can analyze statically checked *programming languages* (e.g. Java, Haskell) into several feedback loops (Figure 2):

1. Programmers often think about design details and calculations on a whiteboard or notebook, even before writing code. This *supplementary medium* has its own feedback loop, even though this is often not automatic.
2. The code is written and is then put through the static checker. An error sends the user back to writing code. In the case of success, they are "allowed" to run the program, leading into cycle 3.
    - The execution gulf comprises multiple cycles of the supplementary medium, plus whatever overhead is needed to invoke the compiler (such as build systems).

- ▪ The evaluation gulf is essentially the waiting period before static errors or a successful termination are observed. Hence this is bounded by some function of the length of the code (the same cannot be said for the following cycle 3.)

3. With a runnable program, the user now evaluates the *runtime* behavior. Runtime errors can send the user back to writing code to be checked, or to tweak dynamically loaded data files in a similar cycle.

   - ▪ The execution gulf here may include multiple iterations of cycle 2, each with its own nested cycle 1.
   - ▪ The *evaluation* gulf here is theoretically unbounded; one may have to wait a very long time, or create very specific conditions, to rule out certain bugs (like race conditions) or simply to consider the program as fit for purpose.
   - ▪ By imposing *static checks*, some bugs can be pushed earlier to the evaluation stage of cycle 2, reducing the likely size of the cycle 3 *evaluation* gulf.
   - ▪ On the other hand, this can make it harder to write statically valid code, which may increase the number of level-2 cycles, thus increasing the total *execution* gulf at level 3.
   - ▪ Depending on how these balance out, the total top-level feedback loop may grow longer or shorter.

### 4.1.2 Example: Immediate Feedback

The specific case where the *evaluation* gulf is minimized to be imperceptible is known as *immediate feedback*. Once the user has caused some change to the system, its effects (including errors) are immediately visible. This is a key ingredient of *liveness*, though it is not sufficient on its own. (See *Relations*)

The ease of achieving immediate feedback is obviously constrained by the computational load of the user's effects on the system, and the system's performance on such tasks. However, such "loading time" is not the only way feedback can be delayed: a common situation is where the user has to manually ask for (or "poll") the relevant state of the system after their actions, even if the system finished the task quickly. Here, the feedback could be described as *immediate upon demand* yet not *automatically demanded*. For convenience, we choose to include the latter criterion—automatic demand of result—in our definition of immediate feedback.

In a *REPL* or *shell*, there is a *main* cycle of typing commands and seeing their output, and a *secondary* cycle of typing and checking the command line itself. The output of commands can be immediate, but usually reflects only part of the total effects or even none at all. The user must manually issue further commands afterwards, to check the relevant state bit by bit. The secondary cycle, like all typing, provides immediate feedback in the form of character "echo", but things like syntax errors generally only get reported *after* the entire line is submitted. This evaluation gulf has been reduced in the JavaScript console of web browsers, where the line is "run" in a limited manner on every keystroke. Simple commands without side-effects,[5] such as calls to pure functions, can give instantly previewed results—though partially typed expressions and syntax errors will not trigger previews.

### 4.1.3 Example: Direct Manipulation

Direct manipulation [80] is a special case of an immediate feedback loop. The user sees and interacts with an artefact in a way that is as similar as possible to real life; this typically includes dragging with a cursor or finger in order to physically move a visual item, and is limited by the particular haptic technology in use.

Naturally, because moving real things with one's hands does not involve any waiting for the object to "catch up",[6] direct manipulation is necessarily an immediate-feedback cycle. If, on the other hand, one were to move a figure on screen by typing new coordinates in a text box, then this could still give *immediate feedback* (if the update appears instant and automatic) but would *not* be an example of direct manipulation.

*Spreadsheets* contain a feedback loop for direct manipulation of values and formatting, as in any other WYSIWYG application. Here, there is feedback for every character typed and every change of style. This is not the case in the other loop for formula editing and formula invocation. There, we see a larger execution gulf for designing and typing formulas, where feedback is only given upon committing the formula by pressing enter. This makes it an "immediate feedback" loop only *on-demand*, as defined above.

### 4.1.4 Dimension: Modes of Interaction

The possible interactions in a programming system are typically structured so that interactions, and the associated feedback loops, are only available in certain *modes*. For example, when creating a new project, the user may be able to configure the project through a conversational interface like npm init in modern JavaScript. Such interactions are no longer available once the project is created. This idea of interaction modes goes beyond just programming systems, appearing in software engineering methodologies. In particular, having a separate *implementation* and *maintenance* phase would be an example of two modes.

*Editing vs debugging.* A good example is the distinction between *editing* and *debugging* mode. When debugging a program, the user can modify the program state and get (more) immediate feedback on what individual operations do. In some systems, one can even modify the program itself during debugging. Such feedback loops are not available outside of debugging mode.

---

[5] Of course, these are detected via some conservative over-approximation which excludes expressions that *might* side-effect.

[6] In some situations, such as steering a boat with a rudder, there is a delay between input and effect. But on closer inspection, this delay is between the rudder and the boat; we do not see the hand pass through the wheel like a hologram, followed by the wheel turning a second later. In real life, objects touched directly give immediate feedback; objects controlled further down the line might not!

*Lisp systems* sometimes distinguish between *interpreted* and *compiled* mode. The two modes do not differ just in the efficiency of code execution, but also in the interactions they enable. In the interpreted mode, code can be tested interactively and errors may be corrected during the code execution (see *Error response*). In the compiled mode, the program can only be tested as a whole. The same two modes also exist, for example, in some Haskell systems where the REPL uses an interpreter (GHCi) distinct from the compiler (GHC).

*Jupyter notebooks.* A programming system may also unify modes that are typically distinct. The Jupyter notebook environment does not have a distinct debugging mode; the user runs blocks of code and receives the result. The single mode can be used to quickly try things out, and to generate the final result, partly playing the role of both debugging and editing modes. However, even Jupyter notebooks distinguish between editing a document and running code.

### 4.1.5 Dimension: Abstraction Construction

A necessary activity in programming is going between abstract schemas and concrete instances. Abstractions can be constructed from concrete examples, first principles or through other methods. A part of the process may happen in the programmer's mind: they think of concrete cases and come up with an abstract concept, which they then directly encode in the system. Alternatively, a system can support these different methods directly.

One option is to construct abstractions *from first principles*. Here, the programmer starts by defining an abstract entity such as an interface in object-oriented programming languages. To do this, they have to think what the required abstraction will be (in the mind) and then encode it (in the system).

Another option is to construct abstractions *from concrete cases*. Here, the programmer uses the system to solve one or more concrete problems and, when they are satisfied, the system guides them in creating an abstraction based on their concrete case(s). In a programming language IDE this manifests as the "extract function" refactor, whereas in other systems we see approaches like macro recording.

*Pygmalion.* In Pygmalion [84], all programming is done by manipulating concrete icons that represent concrete things. To create an abstraction, you can use "Remember mode", which records the operations done on icons and makes it possible to bind this recording to a new icon.

*Jupyter notebook.* In Jupyter notebooks, you are inclined to work with concrete things, because you see previews after individual cells. This discourages creating abstractions, because then you would not be able to look inside at such a fine grained level.

*Spreadsheets.* Up until the recent introduction of lambda expressions into Excel, spreadsheets have been relentlessly concrete, without any way to abstract and reuse patterns of computation other than copy-and-paste.

### 4.1.6 Relations

- *Errors* (Section 4.6) A longer evaluation gulf delays the detection of errors. A longer execution gulf can increase the *likelihood* of errors (e.g. writing a lot of code or taking a long time to write it). By turning runtime bugs into statically detected bugs, the combined evaluation gulfs can be reduced.

- *Adoptability* (Section 4.7): The *execution* gulf is concerned with software using and programming in general. The time taken to realize an idea in software is affected by the user's familiarity and the system's *learnability*.
- *Notation* (Section 4.2): Feedback loops are related to *notational structures*. In a system with multiple notations, each notation may have different associated feedback loops. The motto "The thing on the screen is supposed to be the actual thing" [71], adopted in the context of live programming, relates *liveness* to a direct connection between surface and internal notations. The idea is that interactable objects should be equipped with faithful behavior, instead of being intangible shadows cast by the hidden *real* object.

## 4.2 Notation

**How are the different textual / visual programming notations related?**

Programming is always done through some form of notation. We consider notations in the most general sense and include any structured gesture using textual or visual notation. Textual notations primarily include programming languages, but also things like configuration files. Visual notations include graphical programming languages. Other kinds of structured gestures include user interfaces for constructing visual elements used in the system.

### 4.2.1 Dimension: Notational Structure

In practice, most programming systems use multiple notations. Different notations can play different roles in the system. On the one hand, multiple *overlapping notations* can be provided as different ways of programming the same aspects of the system. In this case, each notation may be more suitable to different kinds of users, but may have certain limitations (for example, a visual notation may have a limited expressive power). On the other hand, multiple *complementing notations* may be used as the means for programming different aspects of the system. In this case, programming the system requires using multiple notations, but each notation may be more suitable for the task at hand; think of how HTML describes document structure while JavaScript specifies its behavior.

### 4.2.2 Example: Overlapping Notations

A programming system may provide multiple notations for programming the same aspect of the system. This is typically motivated by an attempt to offer easy ways of completing different tasks: say, a textual notation for defining abstractions and a visual notation for specifying concrete structures. The crucial issue in this kind of arrangement is *synchronizing* the different notations; if they have different characteristics, this may not be a straightforward mapping. For example, source code may allow more elaborate abstraction mechanisms like loops, which will appear as visible repetition in the visual notation. What should such a system do when the user edits a single object that resulted from such repetition? Similarly, textual notation may allow incomplete expressions that do not have an equivalent in the visual notation. For programming systems that use *overlapping notations*, we need to describe how the notations are synchronized.

*Sketch-n-Sketch* [38] employs overlapping notations for creating and editing SVG and HTML documents. The user edits documents in an interface with a split-screen structure that shows source code on the left and displayed visual output on the right. They can edit both of these and changes are propagated to the other view. The code can use abstraction mechanisms (such as functions) which are not completely visible in the visual editor (an issue we return to in *expression geography* below). Sketch-n-Sketch can be seen as an example of a *projectional editor*.[7]

*UML Round-tripping.* Another example of a programming system that utilizes the *overlapping notations* structure are UML design tools that display the program both as source code and as a UML diagram. Edits in one result in automatic update of the other. An example is the Together/J[8] system. To solve the issue of notation synchronization, such systems often need to store additional information in the textual notation, typically using a special kind of code comment. In this example, after the user re-arranges classes in UML diagrams, the new locations need to be updated in the code.

### 4.2.3 Example: Complementing Notations

A programming system may also provide multiple complementing notations for programming different aspects of its world. Again, this is typically motivated by the aim to make specifying certain aspects of programming easier, but it is more suitable when the different aspects can be more clearly separated. The key issue for systems with complementing notations is how the different notations are connected. The user may need to use both notations at the same time, or they may need to progress from one to the next level when solving increasingly complex problems. In the latter case, the learnability of progressing from one level to the next is a major concern.

*Spreadsheets and HyperCard.* In Excel, there are three different complementing notations that allow users to specify aspects of increasing complexity: (i) the visual grid, (ii) formula language and (iii) a macro language such as Visual Basic for Applications. The notations are largely independent and have different degrees of expressive power. Entering values in a grid cannot be used for specifying new computations, but it can be used to adapt or run a computation, for example when entering different alternatives in What-If Scenario Analysis. More complex tasks can be achieved using formulas and macros. A user gradually learns more advanced notations, but experience with a previous notation does not help with mastering the next one. The approach optimizes for easy learnability at one level, but introduces a hurdle for users to surmount in order to get to the second level. The notational structure of *HyperCard* is similar and consists of (i) visual design of cards, (ii) visual programming (via the GUI) with a limited number of operations and (iii) HyperTalk for arbitrary scripting.

---

[7] Technically, traditional projectional editors usually work more directly with the abstract syntax tree of a programming language.

[8] https://www.mindprod.com/jgloss/togetherj.html, accessed 2023-02-05.

*Boxer and Jupyter.* Boxer [16] uses *complementing notations* in that it combines a visual notation (the layout of the document and the boxes of which it consists) with textual notation (the code in the boxes). Here, the textual notation is always nested within the visual. The case of Jupyter notebooks is similar. The document structure is graphical; code and visual outputs are nested as editable cells in the document. This arrangement is common in many other systems such as Flash or Visual Basic, which both combine visual notation with textual code, although one is not nested in the other.

### 4.2.4 Dimensions: Surface Notation and Internal Notation

All programming systems build up structures in memory, which we can consider as an *internal notation* not usually visible to the user. Even though such structures might be revealed in a debugger, they are hidden during normal operation. What the user interacts with instead is the *surface notation*, typically one of text or shapes on a screen. Every interaction with the surface notation alters the internal notation in some way, and the nature of this connection is worth examining in more detail. To do this, we illustrate with a simplified binary choice for the form of these notations.

### 4.2.5 Examples: Implicit vs. Explicit Structure

Let us partition notations into two families. Notations with *implicit structure* present as a sequence of items, such as textual characters or audio signal amplitudes. Those with *explicit structure* present as a tree or graph without an obvious order, such as shapes in a vector graphics editor. These two types of notations can be transformed into each other: the implicit structure contained in a string can be *parsed* into an explicit syntax tree, and an explicit document structure might be *rendered* into a sequence of characters with the same implicit structure.

Now consider an interface to enter a personal name made up of a forename and a surname. For the surface notation, there could be a single text field to hold the names separated with a space; here, the sub-structure is implicit in the string. Alternatively,[9] there could be two fields where the names are entered separately, and their separation is explicit. A similar choice exists for the internal notation built up in memory: is it a single string, or two separate strings?

We can see that these choices give four combinations. More interestingly, they exhibit unique characters owing to two key asymmetries. Firstly, surface notation is mostly used by humans, while the internal notation is mostly used by the computer. Secondly, and most significantly, computer programs can only work with explicit structure, while humans can understand both explicit and implicit structure. Because of the practical consequences of this asymmetry, we will examine the combinations with emphasis on the *internal* notation first.

---

[9] See "The Many Forms of a Single Fact" [47] for the surprising variety of options available.

### 4.2.6 Examples: One String in Memory (Implicitly Structured Internal Notation)

The simplest case here would be with implicit structure in the surface notation, i.e. a single text box for the full name. Edits to the surface are straightforwardly mirrored interally and persisted to disk. This corresponds to *text editing*. We can generalize this to an idea of *sequence editing* if we view the fundamental act as *recording* events to a list over time. For text, these are key presses; for an audio editing interface they would be samples of sound amplitude.

In the other case, with two text boxes, we have *sequence rendering*. The information about the separation of the two strings, present in the interface, is not quite "thrown away" but is made *implicit* as a space character in the string. This combination corresponds to Visual Basic generating code from GUI forms, video editors combining multiple clips and effects into a single stream, and 3D renderers turning scene graphs into pixels. Another example is line-based diff tools, which provide side-by-side views and related interfaces, yet must ultimately forward the user's changes to the underlying text file.

Critically, in both of these cases, a computer program can only manipulate the stored sequences *as* sequences; that is, by inserting, removing, or serially reading. The appealing feature here is that these operations are simple to implement and may be re-usable across many types of sequences. However, any further structure is implicit and, to work with it programmatically, a user must write a program to *parse* it into something explicit. Furthermore, errors introduced at this stage may simply be *recorded* into the sequence, only to be discovered much later in an attempt to use the data.

### 4.2.7 Examples: Two Strings in Memory (Explicitly Structured Internal Notation)

With two text boxes, both notations match, so there is not much work to do. As with sequence editing, edits on the surface can be mirrored to the internal notation. This corresponds to vector graphics editors and 3D modelling tools, as well as *structure editors* for programming languages. For this reason we call this combination *structure editing*.

With a single text field, we have *structure recovery*. Parsing needs to happen each time the input changes. This style is found in the DOM inspector in browser developer tools, where HTML can be edited as text to make changes to the document tree structure. More generally, this is the mode found in compilers and interpreters which accept program source text yet internally work on tree and graph structures. It is also possible to do a sort of structure editing this way, where the experience is made to resemble text editing but the output is explicitly structured.

In both of these cases, in order to write programs to transform, analyze, or otherwise work with the digital artefact the user has created, one can trivially navigate the stored structure instead of parsing it for every use. Parsing is either done away with altogether or is reduced to a transient process that happens during editing; this means errors can be caught at the moment they are introduced instead of remaining latent.

### 4.2.8 Dimension: Primary and Secondary Notations

In practice, most programming systems use multiple notations. Even in systems based on traditional programming languages, the *primary notation* of the language is often supported by *secondary notations* such as annotations encoded in comments and build tool configuration files. However, it is possible for multiple notations to be primary, especially if they are *overlapping* as defined earlier.

*Programming languages.* Programming systems built around traditional programming languages typically have further notations or structured gestures associated with them. The primary notation in UNIX is the C programming language. Yet this is enclosed in a programming *system* providing a multi-step mechanism for running C code via the terminal, assisted by secondary notations such as shell scripts. Some programming systems attempt to integrate tools that normally rely on secondary notations into the system itself, reducing the number of secondary notations that the programmer needs to master. For example, in the Smalltalk descendant Pharo, versioning and package management is done from within Pharo, removing the need for secondary notation such as git commands and dependency configuration files.[10]

*Haskell.* In Haskell, the primary notation is the programming language, but there are also a number of secondary notations. Those include package managers (e.g. the cabal.project file) or configuration files for Haskell build tools. More interestingly, there is also an informal mathematical notation associated with Haskell that is used when programmers discuss programs on a whiteboard or in academic publications. The idea of having such a mathematical notation dates back to the *Report on Algol 58* [72], which explicitly defined a "publication language" for "stating and communicating problems" using Greek letters and subscripts.

### 4.2.9 Dimension: Expression Geography

A crucial feature of a notation is the relationship between the structure of the notation and the structure of the behavior it encodes. Most importantly, do *similar expressions* in a particular notation represent *similar behavior*?[11] Visual notations may provide a more or less direct mapping. On the one hand, similar-looking code in a block language may mean very different things. On the other hand, similar looking design of two HyperCard cards will result in similar looking cards—the mapping between the notation and the logic is much more direct.

*C/C++ expression language.* In textual notations, this may easily not be the case. Consider the two C conditionals:

- if (x==1) { ... } evaluates the Boolean expression x==1 to determine whether x equals 1, running the code block if the condition holds.

---

[10] The tool for versioning and package management in Pharo can still be seen as an *internal* domain-specific language and thus as a secondary notation, but its basic structure is *shared* with other notations in the Pharo system.

[11] See Basman's [2] similar discussion of "density".

- if (x=1) { ... } *assigns* 1 to the variable x. In C, assignment is an expression *returning* the assigned value, so the result 1 is interpreted as true and the block of code is *always* executed.

A notation can be designed to map better to the logic behind it, for example, by requiring the user to write 1==x. This solves the above problem as 1 is a literal rather than a variable, so it cannot be assigned to (1=x is a compile error).

### 4.2.10 Dimension: Uniformity of Notations

One common concern with notations is the extent to which they are uniform. A uniform notation can express a wide range of things using just a small number of concepts. The primary example here is S-expressions from Lisp. An S-expression is either an atom or a pair of S-expressions written (s1 . s2). By convention, an S-expression (s1 . (s2 . (s3 . nil))) represents a list, written as (s1 s2 s3). In Lisp, uniformity of notations is closely linked to uniformity of representation.[12] In the idealized model of LISP 1.5, the data structures represented by an S-expression are what exists in memory. In real-world Lisp systems, the representation in memory is more complex. A programming system can also take a very different approach and fully separate the notation from the in-memory representation.

*Lisp systems.* In Lisp, source code is represented in memory as S-expressions, which can be manipulated by Lisp primitives. In addition, Lisp systems have robust macro processing as part of their semantics: expanding a macro revises the list structure of the code that uses the macro. Combining these makes it possible to define extensions to the system in Lisp, with syntax indistinguishable from Lisp. Moreover, it is possible to write a program that constructs another Lisp program and not only run it interpretively (using the eval function) but compile it at runtime (using the compile function) and execute it. Many domain-specific languages, as well as prototypes of new programming languages (such as Scheme), were implemented this way. Lisp the language is, in this sense, a "programmable programming language". [21, 19]

### 4.2.11 References

*Cognitive Dimensions of Notation* [34] provide a comprehensive framework for analysing individual notations, while our focus here is on how multiple notations are related and how they are structured. It is worth noting that the Cognitive Dimensions also define *secondary notation*, but in a different sense to ours. For them, secondary notation refers to whether a notation allows including redundant information such as color or comments for readability purposes.

The importance of notations in the practice of science, more generally, has been studied by [49] as "paper tools". These are formula-like entities which can be manipulated by humans in lieu of experimentation, such as the aforementioned mathematical notation in Haskell: a "paper tool" for experimentation on a whiteboard. Program-ming notations are similar, but they are a way of communicating with a machine; the experimentation does not happen on paper alone.

---

[12] Notations generally are closely linked to representation in that the notation may mirror the structures used for program representation. Basman et al. [4] refer to this as a distinction between "dead" notation and "live" representation forms).

### 4.2.12 Relations

- *Interaction* (Section 4.1): The feedback loops that exist in a programming system are typically associated with individual notations. Different notations may also have different feedback loops.
- *Adoptability* (Section 4.7): Notational structure can affect learnability. In particular, complementing notations may require (possibly different) users to master multiple notations. Overlapping notations may improve learnability by allowing the user to edit the program in one way (perhaps visually) and see the effect in the other notation (such as code.)
- *Errors* (Section 4.6). A process that merely records user actions in a sequence (such as text editing) will, in particular, record any *errors* the user makes and defer their handling to later use of the data, keeping the errors *latent*. A process which instead treats user actions as edits to a structure, with constraints and correctness rules, will be able to catch errors at the moment they are introduced and ensure the data coming out is error-free.

## 4.3 Conceptual Structure

**How is meaning constructed? How are internal and external incentives balanced?**

### 4.3.1 Dimension: Conceptual Integrity vs. Openness
The evolution of programming systems has led away from *conceptual integrity* towards an intricate ecosystem of specialized technologies and industry standards. Any attempt to unify parts of this ecosystem into a coherent whole will create *incompatibility* with the remaining parts, which becomes a major barrier to adoption. Designers seeking adoption are pushed to focus on localized incremental improvements that stay within the boundaries established by existing practice. This creates a tension between how highly they can afford to value conceptual elegance, and how open they are to the pressures imposed by society. We will turn to both of these opposite ends—*integrity* and *openness*—in more detail.

### 4.3.2 Example: Conceptual Integrity
I will contend that Conceptual Integrity is the most important consideration in system design. It is better to have a system omit certain anomalous features and improvements, but to reflect one set of design ideas, than to have one that contains many good but independent and uncoordinated ideas. (Fred Brooks, *Aristocracy, Democracy and System Design* [8])

Conceptual integrity arises not (simply) from one mind or from a small number of agreeing resonant minds, but from sometimes hidden co-authors and the thing designed itself. (Richard Gabriel, *Designed As Designer* [27])

Conceptual integrity strives to reduce complexity at the source; it employs *unified concepts* that may *compose orthogonally* to generate diversity. Perhaps the apotheosis of this approach can be found in early Smalltalk and Lisp machines, which were complete programming systems built around a single language. They incorporated capabilities commonly provided *outside* the programming language by operating systems and databases. Everything was done in one language, and so everything was represented with the datatypes of that language. Likewise the libraries and idioms of the language were applicable in all contexts. Having a *lingua franca* avoided much of the friction and impedance mismatches inherent to multi-language systems. A similar drive exists in the Python programming language, which follows the principle that "There should be one—and preferably only one—obvious way to do it" in order to promote community consensus on a single coherent style.

In addition to Smalltalk and Lisp, many programming languages focus on one kind of data structure [81]:

- In COBOL, data consists of nested records as in a business form.
- In Fortran, data consists of parallel arrays.
- In SQL, data is a set of relations with key constraints.
- In scripting languages like Python, Ruby, and Lua, much data takes the form of string-indexed hash tables.

Finally, many languages are *imperative*, staying close to the hardware model of addressable memory, lightly abstracted into primitive values and references into mutable arrays and structures. On the other hand, *functional* languages hide references and treat everything as immutable structured values. This conceptual simplification benefits certain kinds of programming, but can be counterproductive when an imperative approach is more natural, such as in external input/output.

### 4.3.3 Example: Conceptual Openness

*Perl, contra Python*. In contrast to Python's outlook, Perl proclaims "There is more than one way to do it" and considers itself "the first postmodern programming language" [91]. "Perl doesn't have any agenda at all, other than to be maximally useful to the maximal number of people. To be the duct tape of the Internet, and of everything else." The Perl way is to accept the status quo of evolved chaos and build upon it using duct tape and ingenuity. Taken to the extreme, a programming system becomes no longer a *system*, properly speaking, but rather a *toolkit for improvising* assemblages of *found* software. Perl can be seen as championing the values of *pluralism*, *compatibility*, or *conceptual openness* over conceptual integrity. This philosophy has been called *Postmodern Programming* [66].

*C++, contra Smalltalk*. Another case is that of C++, which added to C the Object-Oriented concepts developed by Smalltalk while remaining 100 % compatible with C, down to the level of ABI and performance. This strategy was enormously successful for adoption, but came with the tradeoff of enormous complexity compared to languages designed from scratch for OO, like Smalltalk, Ruby, and Java.

*Worse, contra Better*. Richard Gabriel first described this dilemma in his influential 1991 essay *Worse is Better* [26] analyzing the defeat of Lisp by UNIX and C. Because UNIX and C were so easy to port to new hardware, they were "the ultimate computer viruses" despite providing only "about 50 %–80 % of what you want from an operating system and programming language". Their conceptual openness meant that they adapted easily to the evolving conditions of the external world. The tradeoff was decreased conceptual integrity, such as the undefined behaviours of C, the junkyard of working directories, and the proliferation of special purpose programming languages to provide a complete development environment.

*UNIX and Files*. Many programming languages and systems impose structure at a "fine granularity": that of individual variables and other data and code structures. Conversely, systems like UNIX and the Web impose fewer restrictions on how programmers represent things. UNIX insists only on a basic infrastructure of "large objects" [45], delegating all fine-grained structure to client programs. This scores many points for conceptual openness. *Files* provide a universal API for reading and writing byte streams, a low-level construct containing so many degrees of freedom that it can support a wide variety of formats and ecosystems. *Processes* similarly provide a thin abstraction over machine-level memory and processors.

Concepual integrity is necessarily sacrificed for such openness; while "everything is a file" gestures at integrity, in the vein of Smalltalk's "everything is an object", exceptions proliferate. Directories are special kinds of files with special operations, hardware device files require special ioctl operations, and many commands expect files containing newline separators. Additionally, because client programs must supply their *own* structure for fine-grained data and code, they are given little in the way of mutual compatibility. As a result, they tend to evolve into competing silos of duplicated infrastructure [45, 44].

*The Web*. Web HTTP endpoints, meanwhile, have proven to be an even more adaptable and viral abstraction than UNIX files. They operate at a similar level of abstraction as files, but support richer content and encompass internet-wide interactions between autonomous systems. In a sense, HTTP GET and PUT have become the "subroutine calls" of an internet-scale programming system. Perhaps the most salient thing about the Web is that its usefulness came as such a surprise to everyone involved in designing or competing with it. It is likely that, by staying close to the existing practice of transferring files, the Web gained a competitive edge over more ambitious and less familiar hypertext projects like Xanadu [65].

The choice between compatibility and integrity correlates with the personality traits of *pragmatism* and *idealism*. It is pragmatic to accept the status quo of technology and make the best of it. Conversely, idealists are willing to fight convention and risk rejection in order to attain higher goals. We can wonder which came first: the design decision or the personality trait? Do Lisp and Haskell teach people to think more abstractly and coherently, or do they filter for those with a pre-existing condition? Likewise, perhaps introverted developers prefer the cloisters of Smalltalk or Lisp to the adventurous "Wild West" of the Web.

### 4.3.4 Dimension: Composability
In short, *you can get anywhere by putting together a number of smaller steps.* There exist building blocks which span a range of useful combinations. Composability is, in a sense, key to the notion of "programmability" and every programmable system will have some level of composability (e.g. in the scripting language.)

*UNIX* shell commands are a standard example of composability. The base set of primitive commands can be augmented by programming command executables in other languages. Given some primitives, one can "pipe" one's output to another's input (|), sequence (; or &&), select via conditions, and repeat with loop constructs, enabling full imperative programming. Furthermore, command compositions can be packaged into a named "script" which follows the same interface as primitive commands, and named subprograms within a script can also be defined.

In *HyperCard*, the *Authoring Environment* is *non*-composable for programming buttons: there is simply a set of predefined behaviors to choose from. Full scriptability is available only in the *Programming Environment*.

The *Haskell type system*, as well as that of other functional programming languages, exhibits high composability. New types can be defined in terms of existing ones in several ways. These include records, discriminated unions, function types and recursive constructs (e.g. to define a List as either a Nil or a combination of element plus other list.) The C programming language also has some means of composing types that are analogous in some ways, such as structs, unions, enums and indeed even function pointers. For every type, there is also a corresponding "pointer" type. It lacks, however, the recursive constructs permitted in Haskell types.

### 4.3.5 Dimension: Convenience

In short, *you can get to X, Y or Z via one single step*. There are ready-made solutions to specific problems, not necessarily generalizable or composable. Convenience often manifests as "canonical" solutions and utilities in the form of an expansive standard library.

Composability without convenience is a set of atoms or gears; theoretically, anything one wants could be built out of them, but one must do that work. This situation has been criticized as the *Lisp Curse* [92].

Composability *with* convenience is a set of convenient specific tools *along with* enough components to construct new ones. The specific tools themselves could be transparently composed of these building blocks, but this is not essential. They save users the time and effort it would take to "roll their own" solutions to common tasks.

For example, let us turn to a convenience factor of *UNIX* shell commands, having already discussed their composability above. Observe that it would be possible, in principle, to pass all information to a program via standard input. Yet in actual practice, for convenience, there is a standard interface of *command-line arguments* instead, separate from anything the program takes through standard input. Most programming systems similarly exhibit both composability and convenience, providing templates, standard libraries, or otherwise pre-packaged solutions, which can nevertheless be used programmatically as part of larger operations.

### 4.3.6 Dimension: Commonality

Humans can see Arrays, Strings, Dicts and Sets all have a "size", but the software needs to be *told* that they are the "same". Commonality like this can be factored out into an explicit structure (a "Collection" class), analogous to database *normalization*. This way, an entity's size can be queried without reference to its particular details: if c is declared to be a Collection, then one can straightforwardly access c.size.

Alternatively, it can be left implicit. This is less upfront work, but permits instances to *diverge*, analogous to *redundancy* in databases. For example, Arrays and Strings might end up with "length", while Dict and Set call it "size". This means that, to query the size of an entity, it is necessary to perform a case split according to its concrete type, solely to funnel the diverging paths back to the commonality they represent:

```
if (entity is Array or String)  size := entity.length
else if (entity is Dict or Set) size := entity.size
```

### 4.3.7 Examples: Flattening and Factoring

Data structures usually have several "moving parts" that can vary independently. For example, a simple pair of "vehicle type" and "color" might have all combinations of (Car, Van, Train) and (Red, Blue). In this *factored* representation, we can programmatically change the color directly: pair.second = Red or vehicle.colour = Red.

In some contexts, such as class names, a system might only permit such multi-dimensional structure as an *exhaustive enumeration*: RedCar, BlueCar, RedVan, BlueVan, RedTrain, BlueTrain, etc. The system sees a flat list of atoms, even though a human can see the sub-structure encoded in the string. In this world, we cannot simply "change the color to Red" programmatically; we would need to case-split as follows:

```
if (type is BlueCar) type := RedCar
else if (type is BlueVan) type := RedVan
else if (type is BlueTrain) type := RedTrain
...
```

The *commonality* between RedCar, RedVan, BlueCar, and so on has been *flattened*. There is implicit structure here that remains *un-factored*, similar to how numbers can be expressed as singular expressions (16) or as factor products (2,2,2,2). *Factoring* this commonality gives us the original design, where there is a pair of values from different sets.

In *relational databases*, there is an opposition between *normalization* and *redundancy*. In order to fit multi-table data into a *flat* table structure, data needs to be duplicated into redundant copies. When data is *factored* into small tables as much as possible, such that there is only one place each piece of data "lives", the database is in *normal form* or *normalized*. Redundancy is useful for read-only processes, because there is no need to join different tables together based on common keys. Writing, however, becomes risky; in order to modify one thing, it must be synchronized to the multiple places it is stored. This makes highly normalized databases optimized for writes over reads.

### 4.3.8 Remark: The End of History?

Today we live in a highly developed world of software technology. It is estimated that 41,000 person years have been invested into Linux. We describe software development technologies in terms of *stacks* of specialized tools, each of which might capitalize over 100 person-years of development. Programming systems have become programming ecosystems: not designed, but evolved. How can we noticeably improve programming in the face of the overwhelming edifice of existing technology? There are strong incentives to focus on localized incremental improvements that don't cross the established boundaries.

The history of computing is one of cycles of evolution and revolution. Successive cycles were dominated in turn by mainframes, minicomputers, workstations, personal computers, and the Web. Each transition built a whole new technology ecosystem replacing or on top of the previous. The last revolution, the Web, was 25 years ago, with the result that many people have never experienced a disruptive platform transition. Has history stopped, or are we just stuck in a long cycle, with increasingly pent-up pressures for change? If it is the latter, then incompatible ideas now spurned may yet flourish.

### 4.3.9 References
- How to Design a Good API and Why it Matters [6]

### 4.4 Customizability

**Once a program exists in the system, how can it be extended and modified?**

Programming is a gradual process. We start either from nothing, or from an existing program, and gradually extend and refine it until it serves a given purpose. Programs created using different programming systems can be refined to different extents, in different ways, at different stages of their existence.

Consider three examples. First, a program in a conventional programming language like Java can be refined only by modifying its source code. However, you may be able to do so by just adding new code, such as a new interface implementation. Second, a spreadsheet can be modified at any time by modifying the formulas or data it contains. There is no separate programming phase. However, you have to modify the formulas directly in the cell—there is no way of modifying it by specifying a change in a way that is external to the cell. Third, a *self-sustaining* programming system, such as Smalltalk, does not make an explicit distinction between "programming" and "using" phases, and it can be modified and extended via itself. It gives developers the power to experiment with the system and, in principle, replace it with a better system from within.

### 4.4.1 Dimension: Staging of Customization
For systems that distinguish between different stages, such as writing source code versus running a program, customization methods may be different for each stage. In traditional programming languages, customization is done by modifying or adding source code at the programming stage, but there is no (automatically provided) way of customizing the created programs once they are running.

There are a number of interesting questions related to staging of customization. First, what is the notation used for customization? This may be the notation in which a program was initially created, but a system may also use a secondary notation for customization (consider Emacs using Emacs Lisp). For systems with a stage distinction, an important question is whether such changes are *persistent*.

*Smalltalk, Interlisp and similar.* In image-based programming systems, there is generally no strict distinction between stages and so a program can be customized during execution in the same way as during development. The program image includes the programming environment. Users of a program can open this, navigate to a suitable object or a class (which serve as the *addressable extension points*) and modify that. Lisp-based systems such as *Interlisp* follow a similar model. Changes made directly to the image are persistent. The PILOT system for Lisp [89] offers an interactive way of correcting errors when a program fails during execution. Such corrections are then applied to the image and are thus persistent.

*Document Object Model (DOM) and Webstrates*: In the context of Web programming, there is traditionally a stage distinction between programming (writing the code and markup) and running (displaying a page). However, the DOM can also be modified by browser Developer Tools—either manually, by running scripts in a console, or by using a userscript manager such as Greasemonkey. Such changes are not persistent in the default browser state, but are made so by Webstrates [50] which synchronize the DOM between the server and the client. This makes the DOM collaborative, but not (automatically) *live* because of the complexities this implies for event handling.

### 4.4.2 Dimension: Addressing and Externalizability

Programs in all programming systems have a representation that may be exposed through notation such as source code. When customizing a program, an interesting question is whether a customization needs to be done by modifying the original representation, or whether it can be done by *adding* something alongside the original structure.

In order to support customization through addition, a programming system needs a number of characteristics introduced by Basman et al. [4, 5]. First, the system needs to support *addressing*: the ability to refer to a part of the program representation from the outside. Next, *externalizability* means that a piece of addressed state can be exhaustively transferred between the system and the outside world. Finally, *additive authoring* requires that system behaviours can be *changed* by simply *adding* a new expression containing addresses—in other words, anything can be *overriden* without being *erased*. Of particular importance is how addresses are specified and what extension points in the program they can refer to. The system may offer an automatic mechanism that makes certain parts of a program addressable, or this task may be delegated to the programmer.

*Cascading Style Sheets (CSS)*: CSS is a prime example of additive authoring within the Web programming system. It provides rich addressability mechanisms that are partly automatic (when referring to tag names) and partly manual (when using element IDs and class names). Given a web page, it is possible to modify almost any aspect of its appearance by simply *adding* additional rules to a CSS file. The Infusion project [3] offers similar customizability mechanisms, but for behaviour rather than just styling. There is also the recent programming system Varv [7], which embodies additive authoring as a core principle.

*Object Oriented Programming (OOP) and Aspect Oriented Programming (AOP)*: in conventional programming languages, customization is done by modifying the code itself. OOP and AOP make it possible to do so by adding code independently of existing program code. In OOP, this requires manual definition of extension points, i.e. interfaces and abstract methods. Functionality can then be added to a system by defining a new class (although injecting the new class into existing code without modification requires some form of configuration such as a dependency injection container). AOP systems such as AspectJ [48] provides a richer addressing mechanism. In particular, it makes it possible to add functionality to the invocation of a specific method (among other options) by using the *method call pointcut*. This functionality is similar to *advising* in Pilot [89].

### 4.4.3 Dimension: Self-Sustainability

For most programming languages, programming systems, and ordinary software applications, if one wants to customize beyond a certain point, one must go beyond the facilities provided in the system itself. Most programming systems maintain a clear distinction between the *user level*, where the system is used, and *implementation level*, where the source code of the system itself resides. If the user level does not expose control over some property or feature, then one is forced to go to the implementation level. In the common case this will be a completely different language or system, with an associated learning cost. It is also likely to be lower-level—lacking expressive functions, features or abstractions of the user level—which makes for a more tedious programming experience.

It is possible, however, to carefully design systems to expose deeper aspects of their implementation *at the user level*, relaxing the formerly strict division between these levels. For example, in the research system *3-Lisp* [83], ordinarily built-in functions like the conditional if and error handling catch are implemented in 3-Lisp code at the user level.

The degree to which a system's inner workings are accessible to the user level, we call *self-sustainability*. At the maximal degree of this dimension would reside "stem cell"-like systems: those which can be progressively evolved to arbitrary behavior without having to "step outside" of the system to a lower implementation level. In a sense, any difference between these systems would be merely a difference in initial state, since any could be turned into any other.

The other end, of minimal self-sustainability, corresponds to minimal customizability: beyond the transient run-time state changes that make up the user level of any piece of software, the user cannot change anything without dropping down to the means of implementation of the system. This would resemble a traditional end-user "application" focused on a narrow domain with no means to do anything else.

The terms "self-describing" or "self-implementing" have been used for this property, but they can invite confusion: how can a thing describe itself? Instead, a system that can *sustain itself* is an easier concept to grasp. The examples that we see of high self-sustainability all tend to be *Operating System-like*. UNIX is widely established as an operating system, while Smalltalk and Lisp have been branded differently. Nevertheless, all three have shipped as the operating systems of custom hardware, and have similar responsibilities. Specifically: they support the execution of "programs"; they define an interface for accessing and modifying state; they provide standard libraries of common functionality; they define how programs can communicate with each other; they provide a user interface.

*UNIX*: Self-sustainability of UNIX is owed to the combination of two factors. First, the system is implemented in binary files (via ELF[13]) and text files (for configuration). Second, these files are part of the user-facing filesystem, so users can replace and modify parts of the system using UNIX file interfaces.

*Smalltalk and Combined Object Lambda Architectures*: Self-sustainability in Smalltalk is similar to UNIX, but at a finer granularity and with less emphasis on whether things reside in volatile (process) or non-volatile (file) storage. The analogous points are that (1) the system is implemented as objects with methods containing Smalltalk code, and (2) these are modifiable using the class browser and code editor. Combined Object Lambda Architectures, or COLAs [75], are a theoretical system design to improve on the self-sustainability of Smalltalk. This is achieved by generalizing the object model to support relationships beyond classes.

### 4.4.4 References

In addition to the examples discussed above, the proceedings of self-sustaining systems workshops [41, 40] provide numerous examples of systems and languages that are able to bootstrap, implement, modify, and maintain themselves; Gabriel's analysis of programming language revolutions [28] uses *advising* in PILOT, related Lisp mechanisms, and "mixins" in OOP to illustrate the difference between the "languages" and "systems" paradigms.

### 4.4.5 Relations

- *Flattening and factoring* (Section 4.3.7): related in that "customizability" is a form of creating new programs from existing ones; factoring repetitive aspects into a reusable standard component library facilitates the same thing.
- *Interaction* (Section 4.1): this determines whether there are separate stages for running and writing programs and may thus influence what kind of customization is possible.

## 4.5 Complexity

**How does the system structure complexity and what level of detail is required?**

There is a massive gap between the level of detail required by a computer, which executes a sequence of low-level instructions, and the human description of a program in higher-level terms. To bridge this gap, a programming system needs to deal with the complexity inherent in going from a high-level description to low-level instructions.

Ever since the 1940s, programmers have envisioned that "automatic programming" will allow higher-level programming. This did not necessarily mean full automation. In fact, the first "automatic programming" systems referred to higher-level programming languages with a compiler (or an interpreter) that expanded the high-level code into detailed instructions.

---

[13] Executable and Linkable Format.

Most programming systems use *factoring of complexity* and encapsulate some of the details that need to be specified into components that can be reused by the programmer. The details may be encapsulated in a library, or filled in by a compiler or interpreter. Such factoring may also be reflected in the conceptual structure of the system (Section 4.3.7). However, a system may also fully *automate* some aspects of programming. In those cases, a general-purpose algorithm solves a whole class of problems, which then do not need to be coded explicitly. Think of planning the execution of SQL queries, or of the inference engine supporting a logic programming language like Prolog.

### 4.5.1 Remark: Notations

Even when working at a high level, programming involves manipulating some program notation. In high-level functional or imperative programming languages, the programmer writes code that typically has clear operational meaning, even when some of the complexity is relegated to a library implementation or a runtime. When using declarative programming systems like SQL, Prolog or Datalog, the meaning of a program is still unambiguous, but it is not defined operationally—there is a (more or less deterministic) inference engine that solves the problem based on the provided description. Finally, systems based on *programming by example* step even further away from having clear operational meaning—the program may be simply a collection of sample inputs and outputs, from which a (possibly non-deterministic) engine infers the concrete steps of execution.

### 4.5.2 Dimension: Factoring of Complexity

The basic mechanism for dealing with complexity is *factoring* it. Given a program, the more domain-specific aspects of the logic are specified explicitly, whereas the more mundane and technical aspects of the logic are left to a reusable component. Often, this reusable component is just a library. Yet in the case of higher-level programming languages, the reusable component may include a part of a language runtime such as a memory allocator or a garbage collector. In case of declarative languages or programming by example, the reusable component is a general purpose inference engine.

### 4.5.3 Dimension: Level of Automation

Factoring of complexity shields the programmer from some details, but those details still need to be explicitly programmed. Depending on the customizability of the system, this programming may or may not be accessible, but it is always there. For example, a function used in a spreadsheet formula is implemented in the spreadsheet system.

A programming system with higher *level of automation* requires more than simply factoring code into reusable components. It uses a mechanism where some details of the operational meaning of a program are never explicitly specified, but are inferred automatically by the system. This is the approach of *programming by example* and *machine learning*, where behaviour is specified through examples. In some cases, deciding whether a feature is *automation* or merely *factoring of complexity* is less clear: garbage collection can be seen as either a simple case of automation, or a sophisticated case of factoring complexity.

There is also an interesting (and perhaps inevitable) trade-off. The higher the level of automation, the less explicit the operational meaning of a program. This has a wide range of implications. Smaragdakis [82] notes, for example, that this means the implementation can significantly change the performance of a program.

### 4.5.4 Example: Domain-Specific Languages

Domain-specific languages [22] provide an example of factoring of complexity that does not involve automation. In this case, programming is done at two levels. At the lower level, an (often more experienced) programmer develops a domain-specific language, which lets a (typically less experienced) programmer easily solve problems in a particular domain: say, modelling of financial contracts, or specifying interactive user interfaces.

The domain-specific language provides primitives that can be composed, but each primitive and each form of composition has explicitly programmed and unambiguous operational meaning. The user of the domain-specific language can think in the higher-level concepts it provides, and this conceptual structure can be analysed using the dimensions in Section~4.3. As long as these concepts are clear, the user does not need to be concerned with the details of how exactly the resulting programs run.

### 4.5.5 Example: Programming by Example

An interesting case of automation is *programming by example* [58]. In this case, the user does not provide even a declarative specification of the program behavior, but instead specifies sample inputs and outputs. A more or less sophisticated algorithm then attempts to infer the relationship between the inputs and the outputs. This may, for example, be done through program synthesis where an algorithm composes a transformation using a (small) number of pre-defined operations. Programming by example is often very accessible and has been used in spreadsheet applications [35].

### 4.5.6 Example: Next-level Automation

Throughout history, programmers have always hoped for the next level of "automatic programming". As observed by Parnas [70], "automatic programming has always been a euphemism for programming in a higher-level language than was then available to the programmer".

We may speculate whether Deep Learning will enable the next step of automation. However, this would not be different in principle from existing developments. We can see any level of automation as using *artificial intelligence* methods. This is the case for declarative languages or constraint-based languages—where the inference engine implements a traditional AI method (GOFAI, i.e., Good Old Fashioned AI).

### 4.5.7 Relations

- *Conceptual structure* (Section 4.3): In many cases, the factoring of complexity follows the conceptual structure of the programming system.
- *Flattening and factoring* (Section 4.3.7: One typically automates the thing at the lowest level in one's factoring (by making the lowest level a thing that exists outside of the program—in a system or a library)

## 4.6 Errors

**What does the system consider to be an *error*? How are they prevented and handled?**

A computer system is not aware of human intentions. There will always be human mistakes that the system cannot recognize as errors. Despite this, there are many that it *can* recognize, and its design will determine *which* human mistakes can become detectable program errors. This revolves around several questions: What can cause an error? Which ones can be prevented from happening? How should the system react to errors?

Following the standard literature on errors [77], we distinguish four kinds of errors: slips, lapses, mistakes and failures. A *slip* is an error caused by transient human attention failure, such as a typo in the source code. A *lapse* is similar but caused by memory failure, such as an incorrectly remembered method name. A *mistake* is a logical error such as bad design of an algorithm. Finally, a *failure* is a system error caused by the system itself that the programmer has no control over, e.g. a hardware or a virtual machine failure.

### 4.6.1 Dimension: Error Detection

Errors can be identified in any of the *feedback loops* that the system implements. This can be done either by a human or the system itself, depending on the nature of the feedback loop.

Consider three examples. First, in live programming systems, the programmer immediately sees the result of their code changes. Error detection is done by a human and the system can assist this by visualizing as many consequences of a code change as possible. Second, in a system with a static checking feedback loop (such as syntax checks, static type systems), potential errors are reported as the result of the analysis. Third, errors can be detected when the developed software is run, either when it is tested by the programmer (manually or through automated testing) or when it is run by a user.

Error detection in different feedback loops is suitable for detecting different kinds of errors. Many slips and lapses can be detected by the static checking feedback loop, although this is not always the case. For example, consider a "compact" *expression geography* where small changes in code may result in large changes of behaviour. This makes it easier for slips and lapses to produce hard to detect errors. Mistakes are easier to detect through a live feedback loop, but they can also be partly detected by more advanced static checking.

### 4.6.2 Example: Static Typing

In statically typed programming languages like Haskell and Java, types are used to capture some information about the intent of the programmer. The type checker ensures code matches the lightweight specification given using types. In such systems, types and implementation serve as two descriptions of programmer's intent that need to align; what varies is the extent to which types can capture intent and the way in which the two are constructed; that is, which of the two comes first.

### 4.6.3 Examples: TDD, REPL and Live Coding

Whereas static typing aims to detect errors without executing code, approaches based on immediate feedback typically aim to execute (a portion of) the code and let the programmer see the error immediately. This can be done in a variety of ways.

In case of *test-driven development*, tests play the role of specification (much like types) against which the implementation is checked. Such systems may provide more or less immediate feedback, depending on when tests are executed (automatically in the background, or manually). Systems equipped with a read-eval-print loop (REPL) let programmers run code on-the-fly and inspect results. For successful error detection, the results need to be easily observable: a printed output is more helpful than a hidden change of system state. Finally, in live coding systems, code is executed immediately and the programmer's ability to recognize errors depends on the extent to which the system state is observable. In live coded music, for example, you *hear* that your code is not what you wanted, providing an easy-to-use immediate error detection mechanism.

### 4.6.4 Remark: Eliminating Latent Errors

A common aim of error detection is to prevent *latent errors*, i.e. errors that occured at some *earlier* point during execution, but only manifest themselves through an unexpected behaviour later on. For example, we might dereference the wrong memory address and store a junk value to a database; we will only find out upon accessing the database. Latent errors can be prevented differently in different feedback loops. In a live feedback loop, this can be done by visualizing effects that would normally remain hidden. When running software, latent errors can be prevented through a mechanism that detects errors as early as possible (e.g. initializing pointers to null and stopping if they are dereferenced.)

*Elm and time-travel debugging.* One notable mechanism for identifying latent errors is the concept of *time-travel debugging* popularized by the Elm programming language. In time-travel debugging, the programmer is able to step back through time and see what execution steps were taken prior to a certain point. This makes it possible to break execution when a latent error manifests, but then retrace the execution back to the actual source of the error.

### 4.6.5 Dimension: Error Response

When an error is detected, there are a number of typical ways in which the system can respond. The following applies to systems that provide some kind of error detection during execution.

- It may attempt to automatically recover from the error as best as possible. This may be feasible for simpler errors (slips and lapses), but also for certain mistakes (a mistake in an algorithm's concurrency logic may often be resolved by restarting the code.)
- It may proceed as if the error did not happen. This can eliminate expensive checks, but may lead to latent errors later.
- It may ask a human how to resolve the issue. This can be done interactively, by entering into a mode where the code can be corrected, or non-interactively by stopping the system.

Orthogonally to the above options, a system may also have a way to recover from latent errors by tracing back through the execution in order to find the root cause. It may also have a mechanism for undoing all actions that occurred in the meantime, e.g. through transactional processing.

*Interlisp and Do What I Mean (DWIM).* Interlisp's DWIM facility attempts to automatically correct slips and lapses, especially misspellings and unbalanced parentheses. When Interlisp encounters an error, such as a reference to an undefined symbol, it invokes DWIM. In this case, DWIM then searches for similarly named symbols frequently used by the current user. If it finds one, it invokes the symbol automatically, corrects the source code and notifies the user. In more complex cases where DWIM cannot correct the error automatically, it starts an interaction with the user and lets them correct it manually.

### 4.6.6 Relations
- *Feedback loops*: Error detection always happens as part of an individual feedback loop. The feedback loops thus determine the structure at which error detection can happen.
- *Automation:* A semi-automatic error recovery system (such as DWIM) implements a form of automation. The concept of antifragile software [64] is a more sophisticated example of error recovery through automation.
- *Expression geography:* In an expression geography where small changes in notation lead to valid but differently behaved programs, a slip or lapse is more likely to lead to an error that is difficult to detect through standard mechanisms.

### 4.6.7 References
The most common error handling mechanism in conventional programming languages is exception handling. The modern form of exception handling has been described by Goodenough [31]; Ryder et al. [79] documents the history and influences of Software Engineering on exception handling. The concept of *antifragile software* [64] goes further by suggesting that software could improve in response to errors. Work on Chaos Engineering [10] is a step in this direction.

Reason [77] analyses errors in the context of human errors and develops a classification of errors that we adopt. In the context of computing, errors or *miscomputation* has been analysed from a philosophical perspective [23, 20]. Notably, attitudes and approaches to errors also differ for different programming subcultures [73].

## 4.7 Adoptability

**How does the system facilitate or obstruct adoption by both individuals and communities?**

We consider adoption by individuals as the dimension of *Learnability*, and adoption by communities as the dimension of *Sociability*.

### 4.7.1 Dimension: Learnability

Mainstream software development technologies require substantial effort to learn. Systems can be made easier to learn in several ways:

- Specializing to a specific application domain.
- Specializing to simple small-scale needs.
- Leveraging the background knowledge, skills, and terminologies of specific communities.
- Supporting learning with staged levels of complexity and assistive development tools [24]. Better *Feedback Loops* can help (Section 4.1).
- Collapsing heterogeneous technology stacks into simpler unified systems. This relates to the dimensions under *Conceptual Structure* (Section 4.3).

FORTRAN was a breakthrough in programming because it specialized to scientific computing and leveraged the background knowledge of scientists about mathematical formulas. COBOL instead specialized to business data processing and embraced the business community by eschewing mathematics in favor of plain English.

LOGO was the first language explicitly designed for teaching children. Later BASIC and Pascal were designed for teaching then-standard programming concepts at the University level. BASIC and Pascal had second careers on micropocessors in the 90's. These microprocessor programming systems were notable for being complete solutions integrating everything necessary, and so became home schools for a generation of programmers. More recently languages like Racket, Pyret, and Grace have supported learning by revealing progressive levels of complexity in stages. Scratch returned to Logo's vision of teaching children with a graphical programming environment emphasizing playfulness rather than generality.

Some programming languages have consciously prioritized the programmer's experience of learning and using them. Ruby calls itself *a programmer's best friend* by focusing on simplicity and elegance. Elm targets the more specialized but still fairly broad domain of web applications while focusing on simplicity and programmer-friendliness. It forgoes capabilities that would lead to run-time crashes. It also tries hard to make error messages clear and actionable.

If we look beyond programming languages *per se*, we find programmable systems with better learnability. The best example is spreadsheets, which offer a specialized computing environment that is simpler and more intuitive. The visual metaphor of a grid leverages human perceptual skills. Moving all programming into declarative formulas and attributes greatly simplifies both creation and understanding. Research on Live Programming [36, 90] has sought to incorporate these benefits into general purpose programming, but with limited success to date.

HyperCard and Flash were both programming systems that found widespread adoption by non-experts. Like spreadsheets they had an organizing visual metaphor (cards and timelines respectively). They both made it easy for beginners to get started. Hypercard had layers of complexity intended to facilitate gradual mastery.

Smalltalk and Lisp machines were complex but unified. After overcoming the initial learning curve, their environments provided a complete solution for building entire application systems of arbitrary complexity without having to learn other technologies. Boxer [15] is notable for providing a general-purpose programming environment—albeit for small-scale applications—along with an organizing visual metaphor like that of spreadsheets.

### 4.7.2 Dimension: Sociability

Over time, especially in the internet era, social issues have come to dominate programming. Much programming technology is now developed by open-source communities, and all programming technologies are now embedded in social media communities of their users. Therefore, technical decisions that impact sociabilty can be decisive [61]. These include:

- Compatibility: easy integration into standard technology stacks, allowing incremental adoption, and also easy exit if needed. This dynamic was discussed in the classic essay *Worse is Better* [26] about how UNIX beat Lisp.

- Developing with an open source methodology reaps volunteer labor and fosters a user community of enthusiasts. The technical advantages of open source development were first popularized in the essay *The Cathedral and the Bazaar* [76], which observed that "given enough eyeballs, all bugs are shallow". Open source has become the standard for software development tools, even those developed within large corporations.

- Easy sharing of code via package repositories or open exchanges. Prior to the open-source era, commercial marketplaces were important, like VBX components for VisualBasic. Sharing is impeded when languages lack standard libraries, leading to competing dialects, like Scheme [92].

- Dedicated social media communities can be fostered by using them to provide technical support. Volunteer technical support, like volunteer code contributions, can multiply the impact of core developers. In some cases, social media like Stack Exchange has even come to replace documentation.

One could argue that sociabilty is not purely a *technical* dimension, as it includes aspects of product management. Rather, we believe that sociability is a pervasive cross-cutting concern that cannot be *separated* from the technical.

The tenor of the online community around a programming system can be its most public attribute. Even before social media, Flash developed a vibrant community of amateurs sharing code and tips. The Elm language invested much effort in creating a welcoming community from the outset [13]. Attempts to reform older communities have introduced Codes of Conduct, but not without controversy.

On the other hand, a cloistered community that turns its back on the wider world can give its members strong feelings of belonging and purpose. Examples are Smalltalk,

Racket, Clojure, and Haskell. These communities bear some resemblance to cults, with guru-like leaders, and fierce group cohesion.

The economic sustainability of a programming system can be even more important than strictly social and technical issues. Adopting a technology is a costly investment in terms of time, money, and foregone opportunities. Everyone feels safer investing in a technology backed by large corporations that are not going away, or in technologies that have such widespread adoption that they are guaranteed to persist. A vibrant and mature open-source community backing a technology also makes it safer.

Unfortunately, sociability is often in conflict with learnability. Compatibility leads to ever increasing historical baggage for new learners to master. Large internet corporations have invested mainly in technologies relevant to their expert staff and high-end needs. Open-source communities have mainly flourished around technologies for expert programmers "scratching their own itch". While there has been a flow of venture funding into "no-code" and "low-code" programming systems, it is not clear how they can become economically and socially sustainable. By and large, the internet era has seen the ascendancy of expert programmers and the eclipsing of programming systems for "the rest of us".

## 5   Evaluation

The technical dimensions should be evaluated on the basis of how useful they are for designing and analysing programming systems. To that end, this section demonstrates two uses of the framework. First, we use the dimensions to analyze the recent programming system Dark [56], explaining how it relates to past work and how it contributes to the state of the art. Second, we use technical dimensions to identify a new unexplored point in the design space of programming systems and envision a new design that could emerge from the analysis.

### 5.1  Evaluating the Dark Programming System

Dark is a programming system for building "serverless backends", i.e. services that are used by web and mobile applications. It aims to make building such services easier by "removing accidental complexity"[14] resulting from the large number of systems typically involved in their deployment and operation. This includes infrastructure for orchestration, scaling, logging, monitoring and versioning. Dark provides integrated tooling (Figure 3) for development and is described as *deployless*, meaning that deploying code to production is instantaneous.

Dark illustrates the need for the broader perspective of programming systems. Of course, it contains a programming language, which is inspired by OCaml and F#. But Dark's distinguishing feature is that it eliminates the many secondary systems needed for deployment of modern cloud-based services. Those exist outside of a

---

[14] https://roadmap.darklang.com/goals-of-dark-v2.html, accessed 2023-02-05.

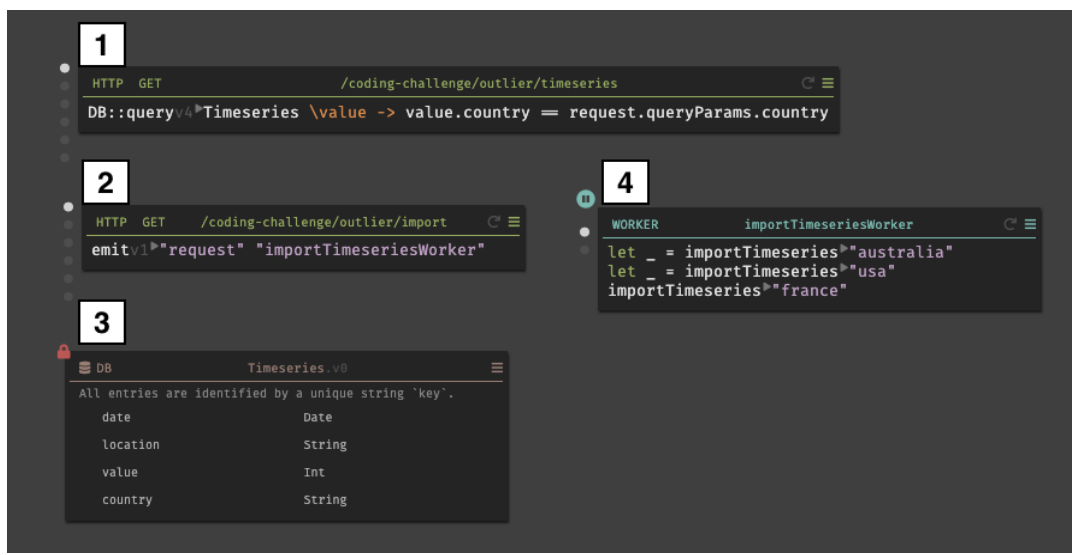

■ **Figure 3**  A simple web service in Dark consisting of two HTTP endpoints (1, 2), a database (3), and a worker (4).

typical programming language, yet form a major part of the complexity of the overall development process.

With technical dimensions, we can go beyond the "sales pitch", look behind the scenes, and better understand the interesting technical aspects of Dark as a programming system. Tables 4 and 5 summarise the more detailed analysis that follows. Two clear benefits of such an analysis are:

1. It provides a list of narrower topics to investigate when examining a programming system such as Dark.
2. It give us a common vocabulary for these topics that can be used to compare Dark with other systems on the same terms.

### 5.1.1 Dimensional Analysis of Dark

**Modes of interaction and feedback loops.**  Conventional *modes of interaction* (4.1.4) include running, editing and debugging. For modern web services, running refers to operation in a cloud-based environment that typically comes with further kinds of feedback (logging and monitoring). The key design decision of Dark is to integrate all these different modes of interaction into a single one. This tight integration allows Dark to provide a more immediate *feedback loop* (4.1.1) where code changes become immediately available not just to the developer, but also to external users. The integrated mode of interaction is reminiscent of the image-based environment in Smalltalk; Dark advances the state of art by using this model in a multi-user, cloud-based context.

**Feedback loops and error response.**  The integration of development and operation also makes it possible to use *errors* occurring during operation to drive development. Specifically, when a Dark service receives a request that is not supported, the user can build a handler [11] to provide a response—taking advantage of the live data that

■ **Table 4** Summary of where Dark lies on its distinguishing dimensions. For brevity, dimensions where Dark does not differ from ordinary programming are omitted.

| Dimension (CLUSTER) | Summary |
| --- | --- |
| **INTERACTION** | |
| Modes of Interaction | Single integrated mode comprises development, debugging and operation ("deployless") |
| Feedback Loops | Code editing is triggered either by user or by unsupported HTTP request and changes are deployed automatically, allowing for *immediate feedback* |
| **ERRORS** | |
| Error Response | When an unsupported HTTP request is received, programmer can write handler code using data from the request in the process |
| **CONCEPTUAL STRUCTURE** | |
| Conceptual Integrity vs. Openness | Abstractions at the domain specific high-level and the functional low-level are both carefully designed for conceptual integrity. |
| Composability | User applications are composed from high-level primitives; the low-level uses composable functional abstractions (records, pipelines). |
| Convenience | Powerful high-level domain-specific abstractions are provided (HTTP, database, workers); core functional libraries exist for the low-level. |

was sent as part of the request. In terms of our dimensions, this is a kind of *error response* (4.6.5) that was pioneered by the PILOT system for Lisp [89]. Dark does this not just to respond to errors, but also as the primary development mechanism, which we might call *Error-Driven Development.* This way, Dark users can construct programs with respect to sample input values.

**Conceptual structure and learnability.**    Dark programs are expressed using high-level concepts that are specific to the domain of server-side web programming: HTTP request handlers, databases, workers and scheduled jobs. These are designed to reduce accidental complexity and aim for high *conceptual integrity* (4.3.1). At the level of code, Dark uses a general-purpose functional language that emphasizes certain concepts, especially records and pipelines. The high-level concepts contribute to *learnability* (4.7.1) of the system, because they are highly domain-specific and will already be familiar to its intended users.

■ **Table 5** Summary of where Dark lies on its distinguishing dimensions. For brevity, dimensions where Dark does not differ from ordinary programming are omitted.

| Dimension (CLUSTER) | Summary |
| --- | --- |
| ADOPTABILITY | |
| Learnability | High-level concepts will be immediately familiar to the target audience; low-level language has the usual learning curve of basic functional programming |
| NOTATION | |
| Notational Structure | Graphical notation for high-level concepts is complemented by structure editor for low-level code |
| Uniformity | Common notational structures are used for database and code, enabling the same editing construct for sequential structures (records, pipelines, tables) |
| COMPLEXITY | |
| Factoring of Complexity | Cloud infrastructure (deployment, orchestration, etc.) is provided by the Dark platform that is invisible to the programmer, but also cannot be modified |
| Level of Automation | Current implementation provides basic infrastructure, but a higher degree of automation in the platform can be provided in the future, e.g. for scalability |
| CUSTOMIZABILITY | |
| Staging of Customization | System can be modified while running and changes are persisted, but they have to be made in the Dark editor, which is distinct from the running service |

**Notational structure and uniformity.** Dark uses a combination of graphical editor and code. The two aspects of the notation follow the *complementing notations* (4.2.1) pattern. The windowed interface is used to work with the high-level concepts and code is used for working with low-level concepts. At the high level, code is structured in freely positionable boxes on a 2D surface. Unlike Boxer [16], these boxes do not nest and the space cannot be used for other content (say, for comments, architectural illustrations or other media). Code at the low level is manipulated using a syntax-aware structure editor, showing inferred types and computed live values for pure functions. It also provides special editing support for records and pipelines, allowing users to add fields and steps respectively.

**Factoring of complexity and automation.** One of the advertised goals of Dark is to remove accidental complexity. This is achieved by collapsing the heterogeneous stack of technologies that are typically required for development, cloud deployment, orchestration and operation. Dark hides this via *factoring of complexity* (4.5.2). The advanced infrastructure is provided by the Dark platform and is hidden from the user. The infrastructure is programmed explicitly and there is no need for sophisticated automation (4.5.3). This factoring of functionality that was previously coded manually follows a similar pattern as the development of garbage collection in high-level programming languages.

**Customizability.**   The Dark platform makes a clear distinction between the platform itself and the user application, so *self-sustainability* (4.4.3) is not an objective. The strict division between the platform and user (related to its aforementioned *factoring of complexity*) means that changes to Dark require modifying the platform source code itself, which is available under a license that solely allows using it for the purpose of contributing. Similarly, applications themselves are developed by modifying and adding code, requiring destructive access to it—so *additive authoring* (4.4.2) is not exhibited at either level. Thanks to the integration of execution and development, persistent changes may be made during execution (c.f. *staging of customization*, 4.4.1) but this is done through the Dark editor, which is separate from the running service.

### 5.1.2  Technical Innovations of Dark

This analysis reveals a number of interesting aspects of the Dark programming system. The first is the tight integration of different *modes of interaction* which collapses a heterogeneous stack of technologies, makes Dark *learnable*, and allows quick feedback from deployed services. The second is the use of *error response* to guide the development of HTTP handlers. Thanks to the technical dimensions framework, each of these can be more precisely described. It is also possible to see how they may be supported in other programming systems. The framework also points to possible alternatives (and perhaps improvements) such as building a more self-sustainable system that has similar characteristics to Dark, but allows greater flexibility in modifying the platform from within itself.

### 5.2  Exploring the Design Space

With a little work, technical dimensions can let us see patterns or gaps in the design space by plotting their values on a simple scatterplot. Here, we will look at two dimensions, *notational diversity*[15] and *self-sustainability*, for the following programming systems: Haskell, Jupyter notebooks, Boxer, HyperCard, the Web, spreadsheets, Lisp, Smalltalk, UNIX, and COLAs.

While our choice to describe dimensions as qualitative concepts was necessary for coming up with them, *some* way of generating numbers is clearly necessary for visualizing their relationships like this. For simplicity,[16] we adopt the following scheme. For each dimension, we distill the main idea into several yes/no questions (Appendix A)

---

[15] This is simply the dimension we named as *uniformity of notations* (4.2.10), but flipped in the opposite direction.

[16] There are undoubtedly many ways to turn our descriptions into various measures, with strengths and weaknesses for different purposes, but this is beyond the scope of the present

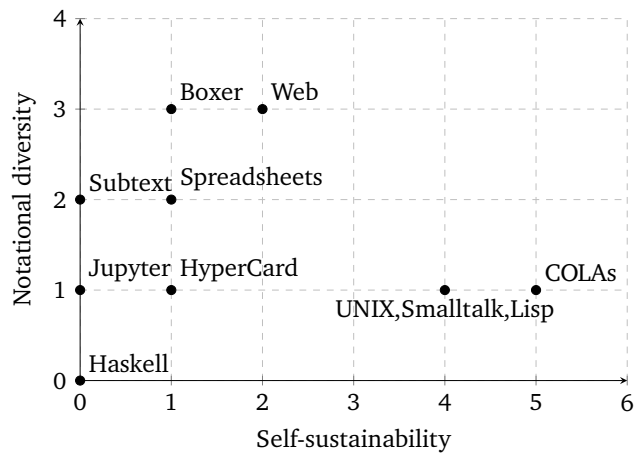

■ **Figure 4** Example programming systems (or system families) measured against *self-sustainability* and *notational diversity*, revealing an absence of systems with a high degree of both.

that capture enough of the distinctions we observe between the systems we wish to plot. Then, for each system, we add up the number of "yes" answers and obtain a plausible score for the dimension.

Figure 4 shows the results we obtained with our sets of questions. It shows that application-focused systems span a range of notational diversity, but only within fairly low self-sustainability. The "OS-likes" (Section 3.2) cluster in an "island" at the right, sharing identical notational diversity and near-identical self-sustainability. There is also a conspicuous blank space at the top-right, representing an unexplored combination of high values on both dimensions. With other pairs of dimensions, we might take this as evidence of an oppositional relationship, such that more of one inherently means less of the other (perhaps looking for a single new dimension that describes this better.) In this case, though, there is no obvious conflict between having many notations and being able to change a system from within. Therefore, we interpret the gap as a new opportunity to try out: combine the self-sustainability of COLAs with the notational diversity of Boxer and Web development. In fact, this is more or less the forthcoming dissertation of the primary author.

## 6 Conclusions

There is a renewed interest in developing new programming systems. Such systems go beyond the simple model of code written in a programming language using a more or less sophisticated text editor. They combine textual and visual notations, create programs through rich graphical interactions, and challenge accepted assumptions

---

paper. Here, we merely wish to demonstrate that such a thing is possible and show what one can do with the results.

about program editing, execution and debugging. Despite the growing number of novel programming systems, it remains difficult to evaluate the design of programming systems and see how they improve over work done in the past. To address the issue, we proposed a framework of "technical dimensions" that captures essential characteristics of programming systems in a qualitative but rigorous way.

The framework of technical dimensions puts the vast variety of programming systems, past and present, on a common footing of commensurability. This is crucial to enable the strengths of each to be identified and, if possible, combined by designers of the next generation of programming systems. As more and more systems are assessed in the framework, a picture of the space of possibilities will gradually emerge. Some regions will be conspicuously empty, indicating unrealized possibilities that could be worth trying. In this way, a domain of "normal science" is created for the design of programming systems.

**Acknowledgements**   We particularly thank Richard Gabriel for shepherding our submission to the 2021 *Pattern Languages of Programming* (PLoP) conference. We thank the participants of the PLoP Writers' Workshop for their feedback, as well as others who have proofread or otherwise given input on the ideas at different stages. These include Luke Church, Filipe Correia, Thomas Green, Brian Hempel, Clemens Klokmose, Geoffery Litt, Mariana Mărășoiu, Stefan Marr, Michael Weiss, and Rebecca and Allen Wirfs-Brock. We also thank the attendees of our Programming 2021 Conversation Starters session and our Programming 2022 tutorial/workshop entitled "Methodology Of Programming Systems" (MOPS).

This work was partially supported by the project of Czech Science Foundation no. 23-06506S.

## A   Making Dimensions Quantitative

To generate numerical co-ordinates for *self-sustainability* and *notational diversity*, we split both dimensions into a small number of yes/no questions and counted the "yes" answers for each system. We came up with the questions informally, with the goal of achieving three things:

1. To capture the basic ideas or features of the dimension
2. To make prior impressions more precise (i.e. to roughly match where we intuitively felt certain key systems fit, but provide precision and possible surprises for systems we were not as confident about.)
3. To be the fewest in number necessary to attain the above

The results of this process were as follows, along with a brief rationale for each question. Afterwards, we will close with some remarks.

## A.1 Self-Sustainability

Questions:

1. *Can you add new items to system namespaces without a restart?* The canonical example of this is in JavaScript, where "built-in" classes like Array or Object can be augmented at will (and destructively modified, but that would be a separate point). Concretely, if a user wishes to make a new sum operation available to all Arrays, they are not *prevented* from straightforwardly adding the method to the Array prototype as if it were just an ordinary object (which it is). Having to re-compile or even restart the system would mean that this cannot be meaningfully achieved from within the system. Conversely, being able to do this means that even "built-in" namespaces are modifiable by ordinary programs, which indicates less of a implementation level vs. user level divide and seems important for self-sustainability.

2. *Can programs generate programs and execute them?* This property, related to "code as data" or the presence of an eval() function, is a key requirement of self-sustainability. Otherwise, re-programming the system, beyond selecting from a predefined list of behaviors, will require editing an external representation and restarting it. If users can type text inside the system then they will be able to write code—yet this code will be inert unless the system can interpret internal data structures as programs and actually execute them.

3. *Are changes persistent enough to encourage indefinite evolution?* If initial tinkering or later progress can be reset by accidentally closing a window, or preserved only through a convoluted process, then this discourages any long-term improvement of a system from within. For example, when developing a JavaScript application with web browser developer tools, it is possible to run arbitrary JavaScript in the console, yet these changes apply only to the running instance. After tinkering in the console with the advantage of concrete system state, one must still go back to the source code file and make the corresponding changes manually. When the page is refreshed to load the updated code, it starts from a fresh initial state. This means it is not worth using the *running* system for any programming beyond tinkering.

4. *Can you reprogram low-level infrastructure within the running system?* This is a hopefully faithful summary of how the COLAs work aims to go beyond Lisp and Smalltalk in this dimension.

5. *Can the user interface be arbitrarily changed from within the system?* Whether classed as "low-level infrastructure" or not, the visual and interactive aspects of a system are a significant part of it. As such, they need to be as open to re-programming as any other part of it to classify as truly self-sustainable.

| Question | Haskell | Jupyter | HyperCard | Subtext | Spreadsheets | Boxer | Web | UNIX | Smalltalk | Lisp | COLAs |
|---|---|---|---|---|---|---|---|---|---|---|---|
| 1 | | | | | | | ✔ | ✔ | ✔ | ✔ | ✔ |
| 2 | | | | | | | ✔ | ✔ | ✔ | ✔ | ✔ |
| 3 | | | ✔ | | ✔ | ✔ | | ✔ | ✔ | ✔ | ✔ |
| 4 | | | | | | | | | | | ✔ |
| 5 | | | | | | | | ✔ | ✔ | ✔ | ✔ |
| Total | 0 | 0 | 1 | 0 | 1 | 1 | 2 | 4 | 4 | 4 | 5 |

## A.2 Notational Diversity

Questions:

1. *Are there multiple syntaxes for textual notation?* Obviously, having more than one textual notation should count for notational diversity. However, for this dimension we want to take into account notations beyond the strictly textual, so we do not want this to be the only relevant question. Ideally, things should be weighted so that having a wide diversity of notations within some *narrow class* is not mistaken for notational diversity in a more global sense. We want to reflect that UNIX, with its vast array of different languages for different situations, can never be as notationally diverse as a system with many languages *and* many graphical notations, for example.

2. *Does the system make use of GUI elements?* This is a focused class of non-textual notations that many of our example systems exhibit.

3. *Is it possible to view and edit data as tree structures?* Tree structures are common in programming, but they are usually worked with as text in some way. A few of our examples provide a graphical notation for this common data structure, so this is one way they can be differentiated from the rest.

4. *Does the system allow freeform arrangement and sizing of data items?* We still felt Boxer and spreadsheets exhibited something not covered by the previous three questions, which is this. Within their respective constraints of rendering trees as nested boxes and single-level grids, they both provide for notational variation that can be useful to the user's context. These systems *could* have decided to keep boxes neatly placed or cells all the same size, but the fact that they allow these to vary scores an additional point for notational diversity.

| Question | Haskell | Jupyter | HyperCard | Subtext | Spreadsheets | Boxer | Web | UNIX | Smalltalk | Lisp | COLAs |
|----------|---------|---------|-----------|---------|--------------|-------|-----|------|-----------|------|-------|
| 1 | | | | | | | ✔ | ✔ | | | ✔ |
| 2 | | ✔ | ✔ | ✔ | ✔ | ✔ | | ✔ | | | |
| 3 | | | | ✔ | | ✔ | ✔ | | | ✔ | |
| 4 | | | | | ✔ | ✔ | | | | | |
| Total | 0 | 1 | 1 | 2 | 2 | 3 | 3 | 1 | 1 | 1 | 1 |

## A.3 Remarks and Future Work

This task of quantifying dimensions forced us to drill down and decide on more crisp definitions of what they should be. We recommend it as a useful exercise even in the absence of a goal like generating a graph.

It is worth clarifying the meaning of what we have done here. It must not be overlooked that this settling down on one particular definition does not replace or obsolete the general qualitative descriptions of the dimensions that we start with. Clearly, there are far too many sources of variation in our process to consider our results here as final, objective, the single correct definition of these dimensions, or anything in this vein. Each of these sources of variation suggests future work for interested parties:

**Quantification goals.** We sought numbers to generate a graph that roughly matched our own intuitive placement of several example systems. In other words, we were trying to make those intuitions more precise along with the dimensions themselves. An entirely different approach would be to have no "anchor" at all, and to take whatever answers a given definition produces as ground truth. However, this would demand more detail for answering questions and generating them in the first place.

**Question generation.** We generated our questions informally and stopped when it seemed like there were enough to make the important distinctions between example points. There is huge room for variation here, though it seems particularly hard to generate questions in any rigorous manner. Perhaps we could take our *self-sustainability* questions to be drawn from a large set of "actions you can perform while the system is running", which could be parametrized more easily. Similarly, our *notational diversity* questions tried to take into account a few classes of notations—a more sophisticated approach might be to just count the notations in a wide range of classes.

**Answering the questions.** We answered our questions by coming to a consensus on what made sense to the three of us. Others may disagree with these answers, and tracing the source of disagreement could yield insights for different questions that both parties would answer identically. Useful information could also be obtained from getting many different people to answer the questions and seeing how much variation there is.

**What is "Lisp", anyway?**  The final major source of variation would be the labels we have assigned to example points. In some cases (Boxer), there really is only one system; in others (spreadsheets) there are several different *products* with different names, yet which are still similar enough to plausibly analyze as the same thing; in still others (Lisp) we're treating a family of related systems as a cohesive point in the design space. It is understandable if some think this elides too many important distinctions. In this case, they could propose splits into different systems or sub-families, or even suggest how these families should be treated as blobs within various sub-spaces.

**About the authors**

**Joel Jakubovic** is Tomas' final-year PhD student interested in overcoming pervasive "premature commitment" in programming. His forthcoming dissertation is about creating programming systems that combine self-sustainability and notational freedom. Send him an e-mail at jdj9@kent.ac.uk and read his blog at programmingmadecomplicated.wordpress.com.

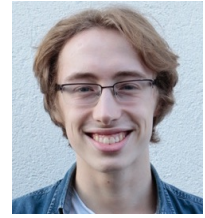

**Jonathan Edwards** is an independent researcher working on drastically simplifying programming. He is known for his Subtext series of programming language experiments and his blog at alarmingdevelopment.org. He has been a researcher at MIT CSAIL and CDG/HARC. He tweets @jonathoda and can be reached at jonathanmedwards@gmail.com.

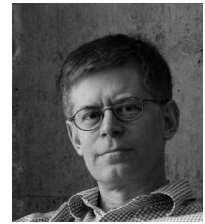

**Tomas Petricek** is an assistant professor at Charles University. He is interested in finding easier and more accessible ways of thinking about programming. To do so, he combines technical work on programming systems and tools with research into history and philosophy of science. His work can be found at tomasp.net and he can be reached at tomas@tomasp.net.

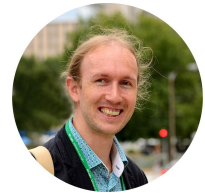